\documentclass[lettersize,journal]{IEEEtran}
\usepackage{booktabs}
\usepackage{multirow}

\usepackage{amsmath,amssymb,amsfonts,amsthm}

\usepackage{textcomp}
\usepackage{url}
\usepackage{bbm}
\usepackage{enumerate}
\usepackage{xcolor}
\usepackage{graphicx}
\usepackage{graphics}
\usepackage{subfigure}
\usepackage{pifont}
\usepackage{lipsum}
\usepackage[T1]{fontenc}
\usepackage{aecompl}
\usepackage{wrapfig}
\usepackage[ruled,linesnumbered]{algorithm2e}
\usepackage{makecell}
\usepackage{multirow}
\usepackage{enumitem}
\usepackage{pifont}
\usepackage{tikz}
\usepackage{tabularx}
\usepackage{booktabs}
\newcommand*\circled[1]{\tikz[baseline=(char.base)]{
            \node[shape=circle,draw,white, fill=black,inner sep=1pt] (char) {#1};}}

\newcommand{\revision}[1]{{\color{black}#1}}

\begin{document}

\title{Resource-Efficient Personal Large Language Models Fine-Tuning with Collaborative Edge Computing}

\author{Shengyuan~Ye, Bei~Ouyang, Tianyi~Qian, Liekang~Zeng, Jingyi~Li, Jiangsu~Du, Xiaowen~Chu, Guoliang~Xing, Xu~Chen% <-this % stops a space

\IEEEcompsocitemizethanks {
\IEEEcompsocthanksitem Shengyuan~Ye, Bei~Ouyang, Tianyi~Qian, Jingyi~Li, Jiangsu~Du are with the School of Computer Science and Engineering, Sun Yat-sen University, Guangzhou 510275, China (e-mail: yeshy8\hspace{0pt}@mail2\hspace{0pt}.sysu\hspace{0pt}.edu\hspace{0pt}.cn; 
ouyb9\hspace{0pt}@mail2\hspace{0pt}.sysu\hspace{0pt}.edu\hspace{0pt}.cn; 
qianty\hspace{0pt}@mail2\hspace{0pt}.sysu\hspace{0pt}.edu\hspace{0pt}.cn; 
lijy573\hspace{0pt}@mail2\hspace{0pt}.sysu\hspace{0pt}.edu\hspace{0pt}.cn; 
dujiangsu\hspace{0pt}@mail\hspace{0pt}.sysu\hspace{0pt}.edu\hspace{0pt}.cn.
\IEEEcompsocthanksitem Liekang~Zeng, and Guoliang~Xing are with the Department of Information Engineering, The Chinese University of Hong Kong, Hong Kong, SAR, China (e-mail: lkzeng@cuhk.edu.hk; glxing@cuhk.edu.hk).
\IEEEcompsocthanksitem Xiaowen~Chu is with The Hong Kong University of Science and Technology (Guangzhou), Guangzhou, China (e-mail: xwchu@ust.hk).
\IEEEcompsocthanksitem Xu~Chen is with the School of Computer Science and Engineering, Sun Yat-sen University, Guangzhou, China and Shenzhen Institute of Artificial Intelligence and Robotics for Society, Shenzhen, China (e-mail: chenxu35@mail.sysu.edu.cn).
}
% \thanks{$*$: Equal contributions. $\dagger$: Corresponding author.}
% <-this % stops a space
% \thanks{Manuscript received April 19, 2005; revised August 26, 2015.}
}

% \markboth{IEEE Transactions on Parallel and Distributed Systems}%

\maketitle

\begin{abstract}

Large language models (LLMs) have enabled transformative applications at the network edge, such as intelligent
personal assistants. However, data privacy and security concerns
necessitate a shift from cloud-centric paradigms to edge-based finetuning for personal LLMs. This transition is significantly hindered by intensive computational requirements and inherent resource scarcity, creating a "resource wall" that compromises training efficiency and feasibility. While current parameter-efficient finetuning (PEFT) and resource management strategies attempt to mitigate these constraints, they remain insufficient for the limited capacities of individual edge devices. To address these challenges, we propose \texttt{PAC+}, a resourceefficient collaborative edge AI framework for in-situ personal LLM fine-tuning. \texttt{PAC+} overcomes the resource bottlenecks through a sophisticated algorithm-system codesign: (1) Algorithmically, \texttt{PAC+} introduces a fine-tuning technique
optimized for parameters, time, and memory. It utilizes Parallel
Adapters to circumvent the need for a full backward pass through
the LLM backbone. Furthermore, an activation cache mechanism
streamlines the process by negating redundant forward passes
across multiple epochs. (2) Systematically, \texttt{PAC+} aggregates proximate edge devices into a collective resource pool, employing hybrid
data and pipeline parallelism to orchestrate distributed training.
By leveraging the activation cache, \texttt{PAC+} enables the exclusive
fine-tuning of Parallel Adapters via data parallelism, effectively
bypassing the backbone’s constraints. Extensive evaluation of the
prototype implementation demonstrates that \texttt{PAC+} significantly
outperforms existing collaborative edge training systems, achieving up to a $9.7\times$ end-to-end speedup. Furthermore, compared to
mainstream LLM fine-tuning algorithms, \texttt{PAC+} reduces memory
footprint by up to $88.16\%$.

\end{abstract}


% \begin{IEEEkeywords}
% Article submission, IEEE, IEEEtran, journal, \LaTeX, paper, template, typesetting.
% \end{IEEEkeywords}

\section{Introduction}
\label{sec:intro}
\begin{figure}[t!]
    \setlength{\abovecaptionskip}{-0.1pt}
    \centering
    \includegraphics[width=\linewidth]{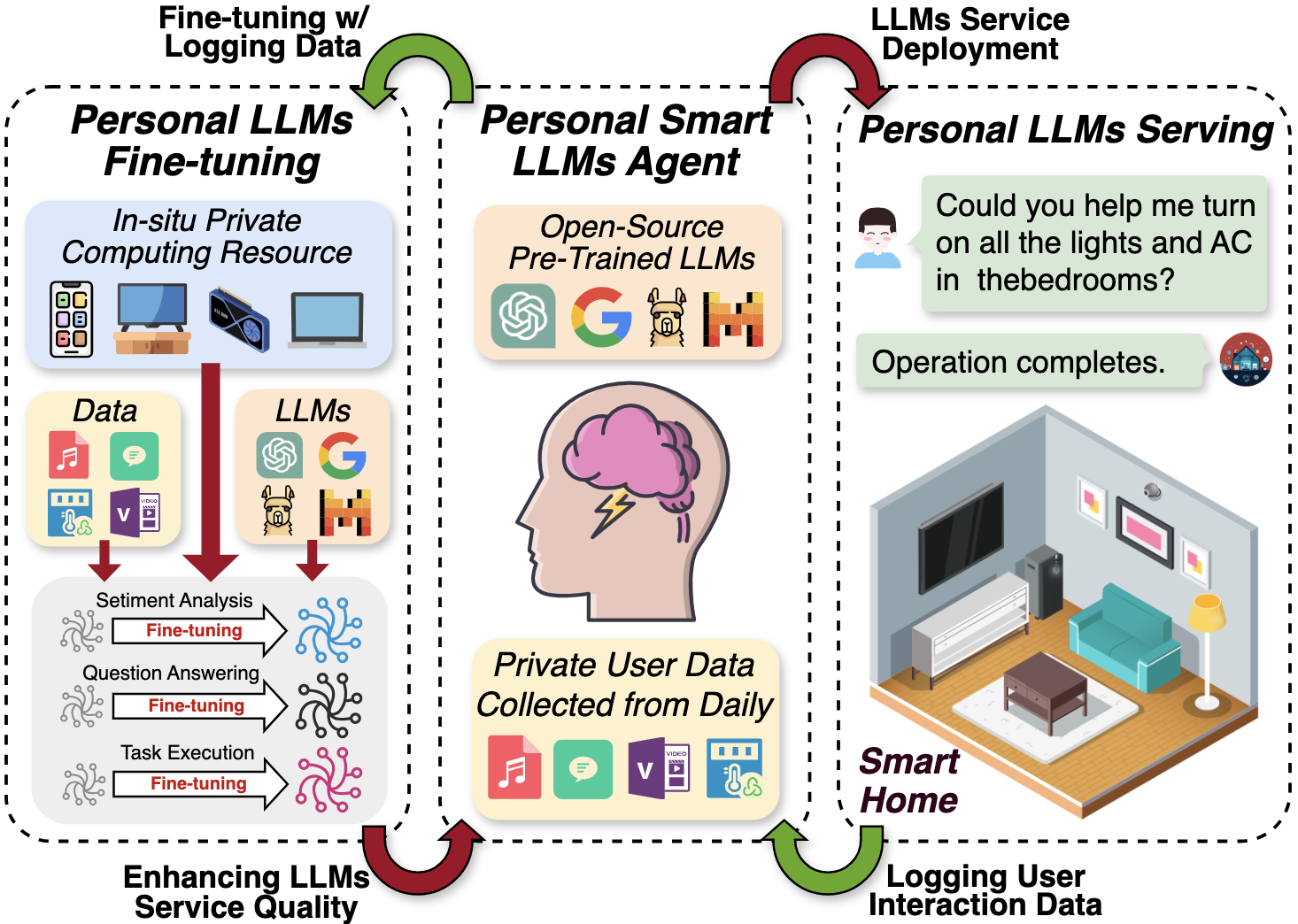}
    \caption{An illustration of hosting personal LLM-based intelligent agents within a smart home.}
    \label{fig:scenario}
    % \vspace{-10pt}
\end{figure}

Large language models (LLMs) \cite{vaswani2017attention, raffel2020exploring, lewis2019bart} have ushered in a revolution in machine intelligence, owing to their exceptional capabilities in a wide range of machine learning tasks. 
While born on datacenter warehouse, LLMs have quickly sunk to edge devices and facilitated a range of intelligent applications at the network edge, such as \textit{intelligent personal assistants} (IPAs) which are software agents that can augment individuals' abilities, complete complicated tasks, and even satisfy emotional needs.
A recent survey \cite{li2024personal} targeting LLM-based IPAs has revealed that over $80\%$ of industry experts believe that, owing to the sensitive and privacy-critical nature of user data, personal LLMs should be fully (or primarily) hosted at the edge in order to enable privacy-preserving model personalization and serving.
Figure \ref{fig:scenario} illustrates the scenario of hosting a personal LLM-based intelligent agent within a smart home.
A personal LLM agent provides users with high-performance, privacy-preserving intelligent services. Meanwhile, the agent also tracks user interactions, learns from experiences, and extracts knowledge to fine-tune the personal LLMs and further enhance the service quality.  

While the serving of LLMs on edge devices has been made feasible through careful engineering	 \cite{guo2023sti, xu2023llmcad, ye2024galaxy}, fine-tuning these models remains significantly challenging due to the resource-intensive nature of LLM training. Towards alleviating the resource challenges, some research works \cite{cai2023efficient, miao2024flexllm} have explored parameter-efficient fine-tuning (PEFT) techniques, such as Adapters \cite{houlsby2019parameter} and LoRA \cite{hu2021lora}, which modify less than $2\%$ of the model's parameters, thereby reducing resource requirements.
Although these techniques are highly parameter-efficient, our analysis observes that they are not resource-efficient enough for edge environments. The inefficiency stems from Adapters and LoRA embedding trainable structures in the LLM backbone, necessitating complete backward passes through the LLMs during backpropagation.
As we will empirically show in \S \ref{sec:background}, fine-tuning a popular LLM of T5-Base (0.25B) by Google with PEFT techniques can only reduce computational overhead by up to $30\%$ compared to full model fine-tuning. 
In practice, fine-tuning the T5-Base on a typical edge device (e.g., NVIDIA Jetson Nano \cite{jetson-nano}) still demands a minimum of $72.6$ minutes per training epoch employing LoRA.
Moreover, on-device fine-tuning is severely hindered by the memory wall of a single device. Predominant techniques require substantial memory to accommodate both model parameters and intermediate results. The observed memory expense for fine-tuning LLMs like T5-Large (0.74B), which exceeds $7.1$ GB with LoRA and $6.8$ GB with 
Adapters, is often unaffordable as typical mobile devices only possess 4-12GB DRAMs in total to run both system software and applications.

Other leading researchers have explored designing sophisticated resource management mechanisms (e.g., CPU-DSP co-execution \cite{xu2022mandheling}, memory budget adapting \cite{gim2022memory, wang2022melon}) to leverage native resources, but are still bottlenecked by the intrinsic resource shortage of single device. To break the resource wall of a single device, we alternatively observe that prevalent edge environments like smart homes usually comprise a group of trusted idle devices beyond a single terminal (e.g., phones and smart-home devices).
These accompanying devices are typically in physical proximity and can be associated as a resource augmentation for in-situ personal LLMs fine-tuning.

As motivated, in this paper, we introduce \textbf{\texttt{PAC+}}, a resource-efficient collaborative edge AI framework for personal LLMs fine-tuning. 
\texttt{PAC+}'s contribution goes beyond merely leveraging distributed edge devices, instead it breaks the resource wall of in-situ personal LLMs fine-tuning with a sophisticated algorithm-system co-design:

\noindent $\bullet$ \textbf{(Algorithm)} 
We evaluate two predominant PEFT techniques, Adapters and LoRA, and reveal that although parameter efficient, these techniques do not achieve sufficient resource efficiency.
\revision{In light of the side-tuning \cite{zhang2020side} techniques, \texttt{PAC+} leverages the existing Parallel Adapters approach \cite{han2024parameter, sung2022lst} to realize not only parameter but also time and memory-efficient personal LLMs fine-tuning.} This technique provides a dedicated gradient "highway" for the trainable parameters, significantly reducing the computational and memory overhead of backpropagation on the LLM backbone.
By employing low-precision quantization on the LLM backbone and tailored initialization methods for the Parallel Adapters, \texttt{PAC+} further minimize the memory and time costs of LLM fine-tuning.
Additionally, our Parallel Adapters stand out from other PEFT techniques by preserving the invariant intermediate activations from the LLM backbone for any given input sequence. 
By reusing these cached activations across multiple epochs, \texttt{PAC+} increases resource efficiency and reduces fine-tuning latency by eliminating repetitive forward propagation through the LLM backbone. 

\noindent $\bullet$ \textbf{(System)} \texttt{PAC+} leverage edge devices in physical proximity and associate them as an edge resource pool for in-situ personal LLMs fine-tuning. Our fine-tuning process can be divided into two phases: (1) For the first epoch, the LLMs backbone, augmented with Parallel Adapters, is fine-tuned across multiple edge devices. To enhance scalability and training throughput, a hybrid parallelism approach that combines the merits of both data and pipeline parallelism is employed by \texttt{PAC+} as a principle to manage collaborative training across multiple edge devices. 
A novel heterogeneity-aware planning algorithm is introduced for multidimensional resources optimization.
(2) In subsequent fine-tuning epochs, the activation cache obviates the need for forward propagation through the LLM backbone, allowing for the exclusive fine-tuning of our Parallel Adapters using data parallelism.

We implement \texttt{PAC+} in realistic testbeds with a cluster of edge devices. 
Extensive evaluations across three LLMs demonstrate that \texttt{PAC+} not only accelerates fine-tuning, achieving up to a $9.7\times$ speedup over existing collaborative edge training systems, but also reduces memory footprint by up to $88.16\%$ compared to mainstream LLM fine-tuning algorithms.

The main contributions are summarized as follows.
\begin{itemize}[leftmargin=*]
    \item We conduct extensive measurement studies on prevalent PEFT techniques, demonstrating their inefficiency, and design a not only parameter but also time and memory efficient LLM fine-tuning technique for resource-limited edge environments.
    \item We analyze and design low-precision quantization for the LLM backbone and tailored initialization methods for the Parallel Adapters, enabling further optimization of both memory and time overhead in LLM fine-tuning.
    \item We design a hybrid parallel fine-tuning architecture to leverage edge devices in physical proximity, complemented by a novel heterogeneity-aware planning algorithm for optimizing multidimensional resources.
    \item We design a system cache mechanism for invariant intermediate activations, eliminating the need for a forward pass through the LLM backbone and enabling exclusive fine-tuning of the Parallel Adapters using data parallelism.
    \item We implement \texttt{PAC+} and evaluate it in realistic edge testbeds. Experimental results show up to $9.7\times$ fine-tuning acceleration and $88.16\%$ memory reduction without sacrificing performance compared to state-of-the-art methods.
\end{itemize}

\begin{figure}[t!]
    \setlength{\abovecaptionskip}{0.1cm}
    \centering
    \includegraphics[width=\linewidth]{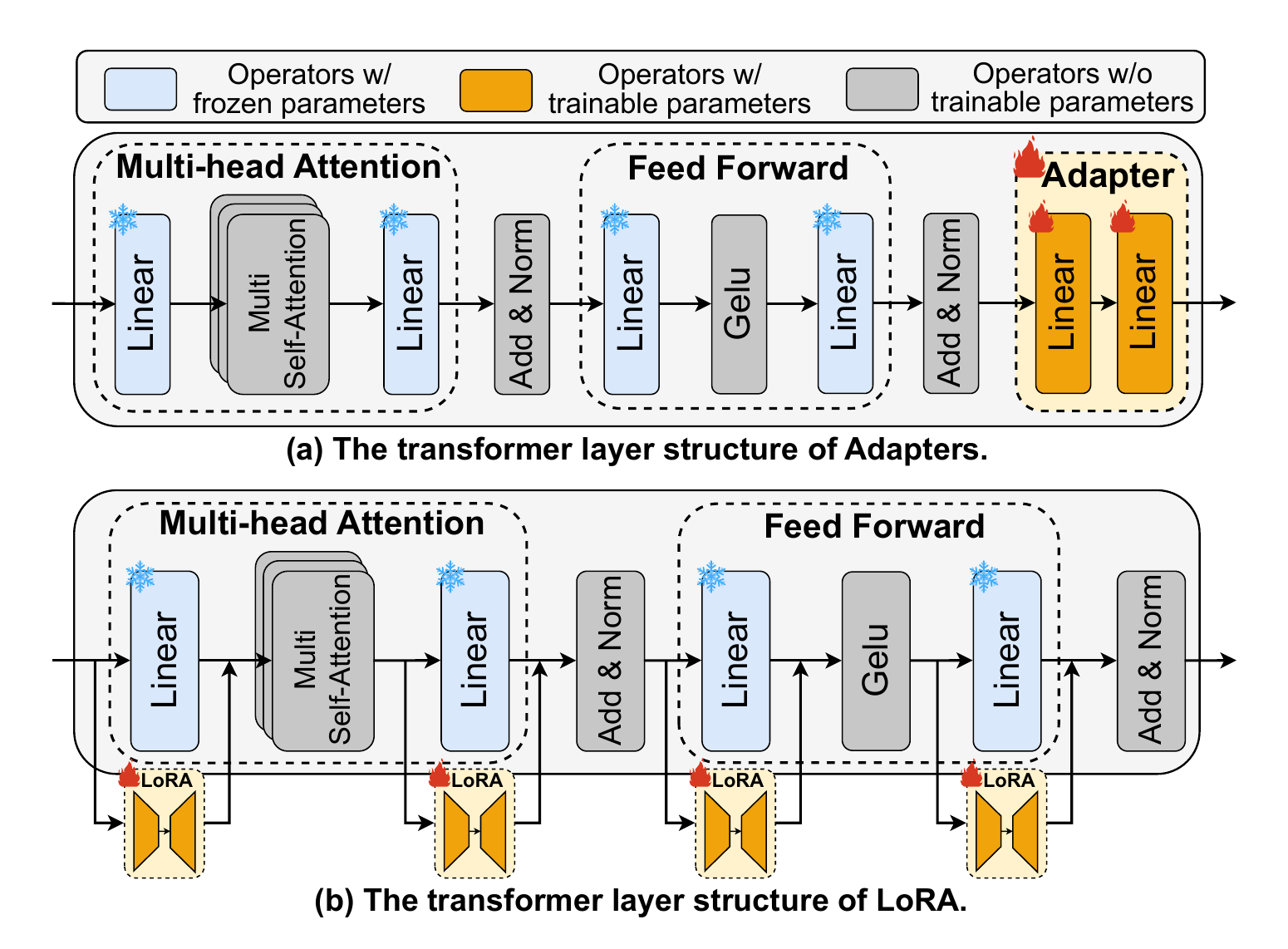}
    \caption{Illustration of the model structures with two PEFT.}
    \label{fig:PEFT}
    \vspace{-15pt}
\end{figure}

\section{MOTIVATION AND PRELIMINARIES}
\label{sec:background}
\subsection{Transformer-Based LLMs and Fine-Tuning}
\textbf{Transformer-Based LLMs.}
Transformer-based LLMs have gained prominence in various language-related applications due to their impressive performance. These models consist of multiple Transformer layers, each comprising two main components: the Multi-head Attention and the Feed Forward block. The Multi-head Attention block utilizes linear layers to generate query (Q), key (K), and value (V) matrices for each attention head, allowing for independent self-attention computations. The outputs of these attention heads are then concatenated and processed through a final linear layer. The Feed Forward block involves two linear operations that expand the hidden dimension size then reduce it back.

\textbf{Personal LLMs Fine-Tuning.} 
The training of LLMs typically consists of two stages: pre-training and fine-tuning. Before being deployed for specific tasks, language models are often pre-trained on extensive text datasets containing vast linguistic data. 
The pre-training process enables the model to acquire a general understanding of linguistic structure and patterns that are widely applicable. 
The fine-tuning adapts the pre-trained model to various, concrete downstream language tasks such as intelligent personal assistants. During actual deployment, the data required for fine-tuning is often generated at the user end, which can carry significant concerns regarding data security and privacy. 
In recent years, in-situ learning on edge devices \cite{patil2022poet, lin2022device, gim2022memory, wang2022melon} has emerged as a promising approach for customizing LLMs while preserving user data fully in-situ.

\begin{figure}[t!]
    \setlength{\abovecaptionskip}{0.1cm}
    \setlength{\belowcaptionskip}{-0.1cm}
    \centering
    \includegraphics[width=\linewidth]{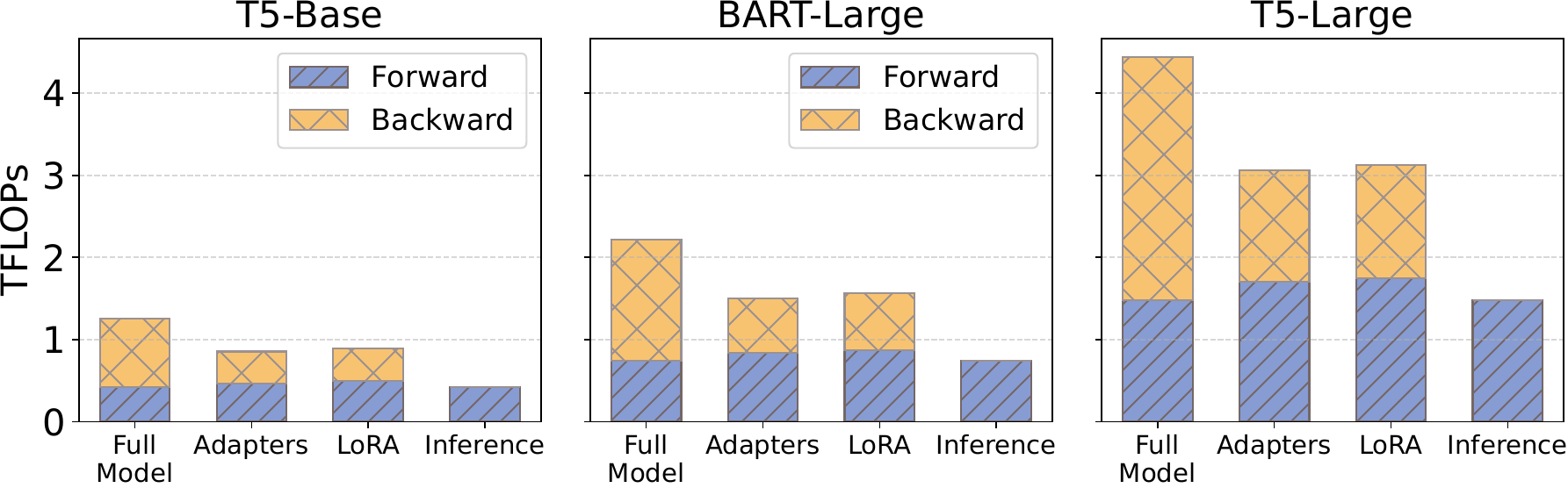}
    \caption{The comparison of floating point of operations (FLOPs). Mini-batch size: 16; sequence length: 128.}
    \label{fig:motivation}
    % \vspace{-5pt}
\end{figure}

% \usepackage{multirow}
\begin{table}[t!]\setlength{\tabcolsep}{5pt}
\setlength{\abovecaptionskip}{0.1cm}
% \small
\caption{The breakdown of memory footprint. "Activations" contain the intermediate results and optimizer states. Model: T5-Large; mini-batch size: 16; sequence length: 128.}
\begin{tabular}{cccccc}
\toprule[1pt]
\multirow{2}{*}{Techniques} & \multirow{2}{*}{\begin{tabular}[c]{@{}c@{}}Trainable\\ Parameters\end{tabular}} & \multicolumn{4}{c}{Memory Footprint (GB)} \\ \cline{3-6} 
                         &                                                                                 & Weights    & Activations    & Gradients    & Total   \\ \hline \hline
Full                     & 737M (100\%)                                                                    & 2.75          & 5.33       & 2.75        & 10.83      \\
Adapters                  & 12M (1.70 \%)                                                                      & 2.80          & 4.04       & 0.05       & 6.89   \\
LoRA                     & 9M (1.26\%)                                                                      & 2.78          & 4.31       &  0.04      &  7.13   \\
Inference                & /                                                                               &  2.75          & /   &  /      &     2.75   \\ \toprule[1pt]
\end{tabular}
\label{tab:memory}
% \vspace{-15pt}
\end{table}

Full model fine-tuning updates all parameters of an LLM for a specific downstream task. However, it is impractical for adapting an LLM to multiple distinct downstream tasks, as each target task would require maintaining a separate LLM with whole parameters.
Some leading researchers have proposed parameter-efficient fine-tuning (PEFT) techniques \cite{lester2021power,houlsby2019parameter,hu2021lora,liu2024tuning} which adapt a small subset of the LLM parameters or a set of newly added parameters for each new task. 
Adapters \cite{houlsby2019parameter} and LoRA \cite{hu2021lora} are two of the most widely used PEFT techniques. Figure \ref{fig:PEFT} illustrates how the transformer layer structure incorporates these two techniques. 
Specifically, adapters are compact bottleneck modules inserted at the end of each transformer layer. 
Similarly, LoRA injects trainable low-rank matrices into a frozen pre-trained model. These decompose the weight matrix parameter updates into two learnable low-rank matrices.
Extensive experiments have demonstrated that these PEFT techniques can achieve performance comparable to full fine-tuning. 

Although these PEFT techniques can greatly reduce the number of trainable parameters (around $98\%$), our analysis has revealed that they do not significantly decrease the computational and memory requirements during training.
Figure \ref{fig:motivation} illustrates the floating point of operations (FLOPs) of different fine-tuning techniques and inference. Adapters and LoRA exhibit a limited reduction in computation (around $30\%$).
Table \ref{tab:memory} summarizes the memory footprint breakdown for T5-Large.
Although Adapters and LoRA minimize the gradient memory footprint by restricting the number of trainable parameters, the memory consumed by activations still constitutes substantial overhead, with a maximum reduction of only $36\%$.
The reason is that both Adapters and LoRA introduce trainable structures within the LLM backbone, such as at the end of each transformer block or as bypasses to linear layers. Computing gradients for trainable parameters via backpropagation involves traversing the LLM backbone, compromising the efficiency of PEFT techniques due to the additional computational overhead and memory required to maintain considerable intermediate activations in LLM backbone.

\begin{figure*}[t!]
    \setlength{\abovecaptionskip}{0.1cm}
    \centering
    \includegraphics[width=\linewidth]{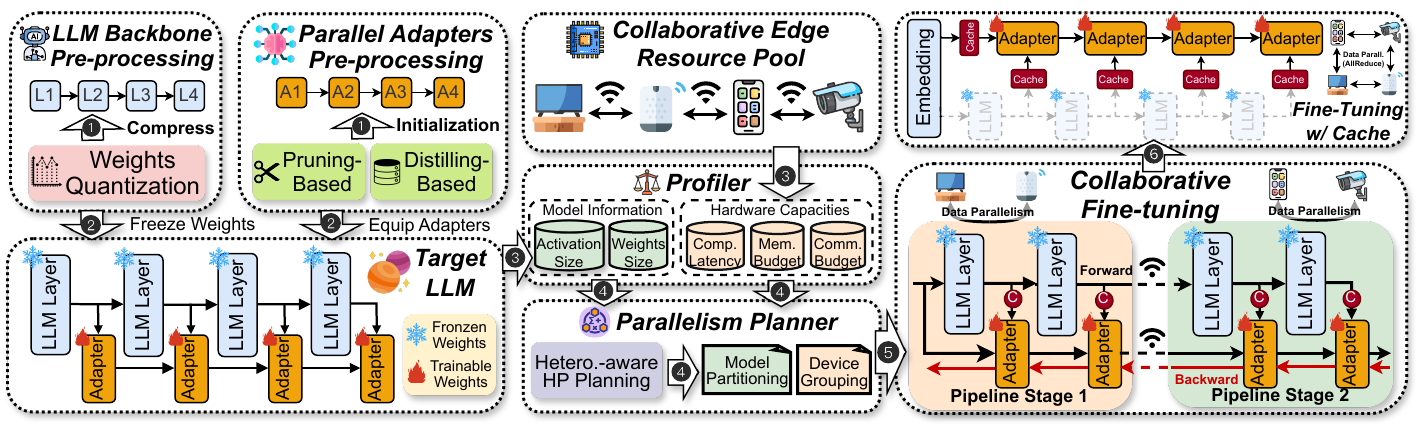}
    \caption{\revision{\texttt{PAC+} workflow.}}
    \label{fig:overview}
    \vspace{-15pt}
\end{figure*}

\subsection{Personal LLMs Fine-Tuning with Resource-Constrained Edge Devices} 
\label{sec:back-personal}
On-device fine-tuning enables leveraging idle resources at the edge while fully preserving user data privacy \cite{patil2022poet, lin2022device, gim2022memory, wang2022melon}. This paradigm is widely adopted in privacy-sensitive edge computing applications. However, the resource-intensive nature of LLMs fine-tuning presents two significant challenges for resource-limited edge devices: \textbf{(1) The computational capabilities of edge devices are constrained.} 
Edge devices often face stark computational constraints compared to the powerful accelerators available in cloud datacenters. The Jetson Nano \cite{jetson-nano}, a specialized platform for edge AI, peaks at a mere $0.47$ TFLOPS, a tiny fraction of the $312$ TFLOPS achievable with NVIDIA's A100 GPU typically found in data centers. Fine-tuning a T5-Base model with Adapters on a single Jetson Nano requires an epoch time of 72.6 minutes, which is $175.5\times$ longer than that in a NVIDIA A100 GPU, showing the fundamental contradiction between intensive LLM fine-tuning workload and constrained on-board resources.
\textbf{(2) On-device fine-tuning is hindered by the memory wall}. As shown in Table \ref{tab:memory}, fine-tuning the T5-Large model incurs a peak memory footprint that is often unaffordable for edge devices. 
For instance, although PEFT techniques such as Adapters and LoRA adjust only approximately $2\%$ of the parameters, they still require substantial memory $6.89$ GB and $7.13$ GB respectively. Compared to full model fine-tuning, which requires over $10$ GB, these techniques reduce memory usage by only $36\%$, often insufficient for typical mobile devices with 4-12 GB DRAM to run system software and applications.

To break the resource wall of a single edge device, in our work, we alternatively observe that prevalent edge scenarios usually comprise a group of trusted idle devices beyond a single terminal. These accompanying devices are typically located in close physical proximity, such as being connected to the same local area network (LAN), and can be utilized as a resource augmentation for in-situ LLMs fine-tuning acceleration. 
While several pioneering research works \cite{ye2024galaxy, wei2024communication} have delved into collaborative edge computing to overcome resource limitations faced by edge devices, the majority of these works primarily focus on LLMs inference.
Other studies \cite{cai2023efficient, xu2024fwdllm} employing federated learning for fine-tuning LLMs with collaborative edge devices primarily address the dissolution of data silos, rather than resource augmentation within LANs.

\subsection{Technical Challenges}
According to the aforementioned analysis and discussion, we can summarize two key challenges inherent to fine-tuning personal LLMs on resource-constrained edge devices:

\textbf{(1) How to boost the time and memory efficiency for edge fine-tuning?} 
The current predominant PEFT techniques like LoRA and Adapters, while parameter-efficient, are not inherently resource-efficient. These approaches still require substantial computational and memory resources for deployment. This poses a fundamental contradiction, as edge devices are typically constrained by cost and power considerations, and thus possess highly limited computational and memory capabilities.
Therefore, we need to exploit a PEFT technique that is inherently more time and memory efficient, and well-suited for edge environments.

\textbf{(2) How to orchestrate resource-efficient collaborative computing across multiple edge devices?} 
Leveraging multiple edge devices for distributed fine-tuning is not trivial. Careful selection of parallel architectures and parallelism planning strategies is necessary to maximize resource utilization and avoid out-of-memory issues.
Furthermore, an algorithm-system co-design approach is required to seamlessly integrate the resource-efficient PEFT techniques with our collaborative edge system. This holistic design is crucial for enabling scalable and high-performance personal LLMs fine-tuning in resource-constrained edge environments.

\vspace{-10pt}
\section{System Overview}
\label{sec:system-design}
\texttt{PAC+} is a resource efficient collaborative framework for personal LLMs fine-tuning across multiple edge devices. Figure \ref{fig:overview} illustrates the \texttt{PAC+} workflow.
\texttt{PAC+} first constructs a lightweight Parallel Adapter for parameter-efficient LLM fine-tuning, developed as a lightweight version of the target transformer-based LLM.
Next, \texttt{PAC+} optimize resource efficiency during fine-tuning by pre-processing both the LLM and Parallel Adapters. This includes quantizing LLM parameters to a lower bit-width representation for reduced memory footprint, and applying initialization techniques such as structural pruning or knowledge distillation to accelerate convergence (Step \circled{1}).
\texttt{PAC+} then integrate the Parallel Adapters module into the LLM, freezing the LLM backbone parameters and fine-tuning only the lightweight adapter for parameter efficient adapting (Step \circled{2}).
\texttt{PAC+} profiler fine-tunes the LLM using a calibration dataset on edge devices to record the runtime profile required for parallelism planning (Step \circled{3}).
\texttt{PAC+} planner then takes the profiling results as input and generates planning configurations, including LLM partitioning points and device grouping strategies. These configurations comprehensively addresses challenges including memory budget, limited communication capacity, and resource heterogeneity (Step {\circled{4}).
The parallel configurations generated by the \texttt{PAC+} planner are then applied to the edge devices, enabling time, memory, and parameter-efficient hybrid data and pipeline parallelism fine-tuning of the target LLM (Step \circled{5}). 
Since the LLM backbone parameters remain fixed, the intermediate activations generated by the backbone model are invariant for a given input sequence. The \texttt{PAC+} maintains a cache of these invariant activations. Through leveraging the cached activations, the efficiency of the fine-tuning process can be accelerated (Step \circled{6}).

\section{Time, Memory and Parameter Efficient Fine-Tuning Algorithm}
\subsection{Fine-Tuning LLMs with Parallel Adapters}
\label{sec:paral-adapter}
\textbf{Observation and Key Insight.} 
As discussed in \S \ref{sec:background}, while techniques such as LoRA \cite{hu2021lora} and Adapters \cite{houlsby2019parameter} reduce the number of parameters that need to be updated during fine-tuning, they do not significantly reduce the computational and memory requirements during the training on edge devices. 
This is because the parameters being updated are still inside the LLM backbone. To calculate the gradients for backpropagation, the full backward passes through the entire pre-trained model are still necessary, as illustrated in Figure \ref{fig:paral-ada}(a) and (b).
In the research field of AI, side-tuning \cite{zhang2020side} is a specialized fine-tuning technique. It adds a trainable side network that runs in parallel to the backbone model, with the side network's representation summed with the backbone's output in the final layer. Crucially, side-tuning only updates the side network, without backpropagating through the backbone model.

\begin{figure}[t!]
    \setlength{\abovecaptionskip}{0.1cm}
    \centering
    \includegraphics[width=0.92\linewidth]{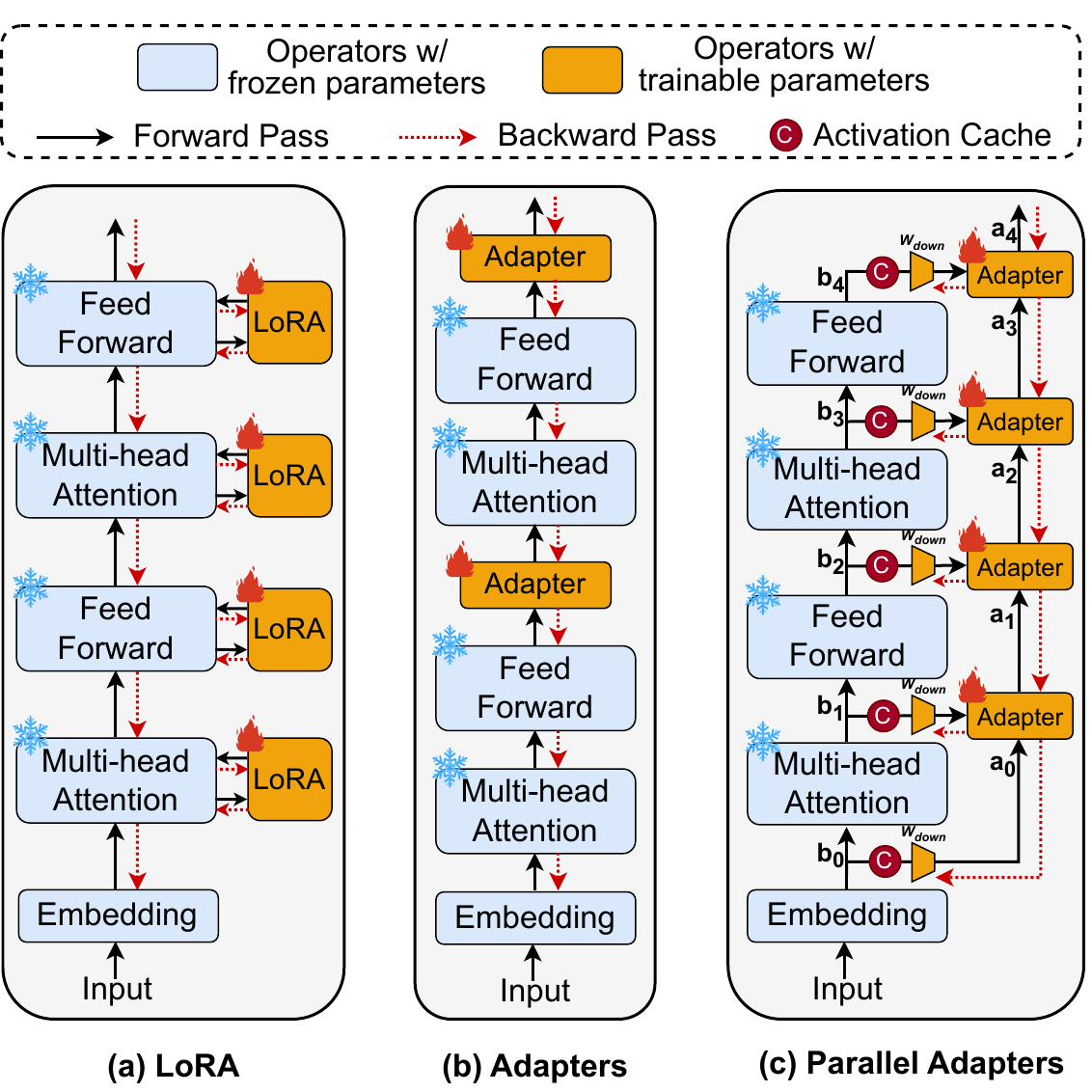}
    \caption{Comparison between LLMs fine-tuning with LoRA, Adapters, and our Parallel Adapters.}
    \label{fig:paral-ada}
    % \vspace{-15pt}
\end{figure}

\textbf{Parallel Adapters Architecture.} 
  \label{sec:activatoin_cache}
In light of side-tuning, we employ a time and memory efficient personal LLMs fine-tuning technique with Parallel Adapters. 
The overall structure is illustrated in Figure \ref{fig:paral-ada}(c). 
Specifically, we decouple conventional Adapters \cite{houlsby2019parameter} from the LLM backbone, avoiding their integration at the end of each transformer layer. Instead, we provide a dedicated parallel highway for our trainable adapters network, which takes intermediate activations from the backbone transformer as input and generates the final predictions. In this way, backpropagation through the LLM backbone is free, reducing memory demands for massive activations and computational burdens, thereby enhancing time and memory efficiency over techniques like Adapters and LoRA.
Observing Figure \ref{fig:paral-ada}(c), the training paradigm of our Parallel Adapters can be viewed as a shift from fine-tuning the original LLM backbone to fine-tuning a lightweight proxy network that operates in parallel with the backbone.
In our design, this lightweight proxy network (i.e., Parallel Adapter) is developed as a lightweight version of the backbone transformer-based LLM \cite{han2024parameter, sung2022lst}. 
Additionally, our adapter module also demonstrates broad compatibility with established LLM fine-tuning architectures, including the commonly used linear layers for upward and downward projections \cite{houlsby2019parameter}.

Figure \ref{fig:pa-detail} provides a detailed illustration of the workflow of our Parallel Adapters architecture.
To ensure the lightweight and resource-efficient nature of our proxy network, the hidden dimension of our Parallel Adapters will be $\frac{1}{r}$ times of the original hidden dimension in backbone.
Consider an LLM backbone with $L$ layers, where the intermediate outputs $\mathrm{b}_1, \mathrm{b}_2, \ldots, \mathrm{b}_L$ each contain $n$ tokens, with a hidden dimensionality of $d$, such that $\mathbf{b}_i \in \mathbb{R}^{n \times d}$.
We denote the embedding input sequence as $\mathrm{b}_0 \in \mathbb{R}^{n \times d}$.
Assuming adapters are inserted after every transformer layer of backbone LLM, Parallel Adapters consist of $L$ adapters, yielding $L$ intermediate outputs $\mathrm{a}_1,\mathrm{a}_2,\ldots\mathrm{a}_L$, $\mathbf{a}_i\in\mathbb{R}^{n\times \frac{d}{r}}$.
To align the dimension of $\mathrm{b}_{i+1}$ with $\mathrm{a}_i$, we learn a linear projection $\mathbf{W}_{down}^i \in \mathbb{R}^d\to\mathbb{R}^{\frac{d}{r}}$ to downsample ($ \times \frac{1}{r}$) the intermediate activations from the LLM backbone, resulting in $\frac{\mathrm{b}_{i+1}}{r} \in \mathbb{R}^{n \times \frac{d}{r}}$.
Next, we combine the downsampled LLM activations $\frac{\mathrm{b}_{i+1}}{r}$ with the intermediate results of the Parallel Adapters $\mathrm{a}_i$ using a learnable parameter $\lambda_i$: $\lambda_i \frac{b_{i+1}}{r} + (1-\lambda_i) a_i$. 
This combined result is then used as the input to the $i$-th adapter, with $\lambda_i$ initialized to 0.5.
At the end of the Parallel Adapters, we learn a linear projection $\mathbf{W}_{up} \in \mathbb{R}^{\frac{d}{r}}\to\mathbb{R}^{d}$ to upsample the final outputs to match the dimension of the original language model head.
Our evaluation in \S \ref{sec:eval} reveals that Parallel Adapters can achieve comparable model performance to mainstream fine-tuning techniques while being more resource-efficient and better suited for resource-constrained edge environments.

\begin{figure}[t!]
    \setlength{\abovecaptionskip}{0.1cm}
    \centering
    \includegraphics[width=0.92\linewidth]{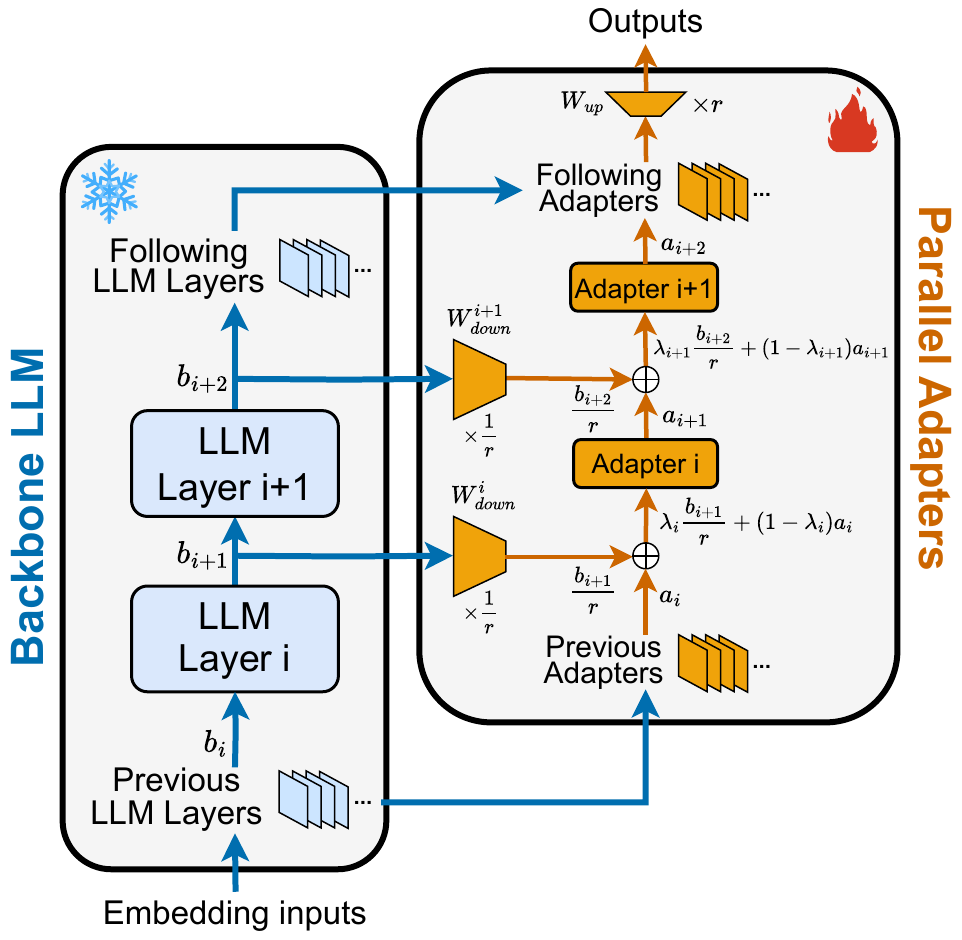}
    \caption{Detailed illustration of the Parallel Adapters architecture.}
    \label{fig:pa-detail}
    % \vspace{-15pt}
\end{figure}

\subsection{Activation Cache for Parallel Adapters}
\textbf{Observation and Opportunities.}
Leveraging Parallel Adapters substantially diminishes the computational and memory demands by circumventing backward propagation through the LLM backbone. 
However, for edge environments with limited resources, forward propagation calculations on the backbone of LLMs also require substantial computational resources. 
Figure \ref{fig:motivation} demonstrates that the computational overhead for forward propagation constitutes $54\%$ and $56\%$ of the total overhead when fine-tuning the T5-Large with Adapters and LoRA, respectively.

To minimize the computational demand, we identify two distinct opportunities for utilizing Parallel Adapters in in-situ fine-tuning of LLMs: (1) During the pre-training phase of LLMs, due to the vast volumes of data involved, researchers typically train for only one epoch, meaning each sequence input is processed by the model a single time. 
However, in typical in-situ LLM fine-tuning scenarios, users often utilize small datasets collected from their specific context, repeatedly training the models with these inputs until achieving model convergence.
(2) When employing parallel adapters to fine-tune LLMs, the parameters of the LLM backbone remain fixed.  
Unlike other PEFT techniques, the LLM backbone operates independently of the intermediate outputs generated by Parallel Adapters.
Consequently, for a given input sequence, the activations generated by the LLM backbone are always invariant.

\textbf{Fine-Tuning Parallel Adapters with \texttt{PAC+} Activation Cache.} 
Our key idea leverages the frozen parameters of the backbone model, enabling the caching of activations produced during the forward propagation of the same input sequence, thereby facilitating their reuse across multiple epochs \cite{cai2023efficient}.
As discussed in \S \ref{sec:paral-adapter}, the parallel adapters are a lightweight, separate network that takes the intermediate activations from the backbone transformer as input and generates predictions. 
During the first epoch, when processing a new input sequence, we cache all the input activations required by the Parallel Adapters that are obtained from the LLM backbone, as illustrated in Figure \ref{fig:paral-ada}(c), highlighted by the red circle.
In subsequent fine-tuning epochs using the same input sequence, we can skip the forward propagation through the LLM backbone entirely, since the required activations have already been cached.
The combination of Parallel Adapters and activation caching allows efficient fine-tuning of the LLMs without the need for both forward and backward propagation through the backbone network, thereby (1) significantly accelerating the fine-tuning process and (2) reducing the memory footprint by allowing the release of the memory space occupied by the LLM parameters. 
% benefits that are particularly valuable for resource-constrained edge environments.

\begin{figure}[t!]
    \setlength{\abovecaptionskip}{0.1cm}
    \centering
    \includegraphics[width=\linewidth]{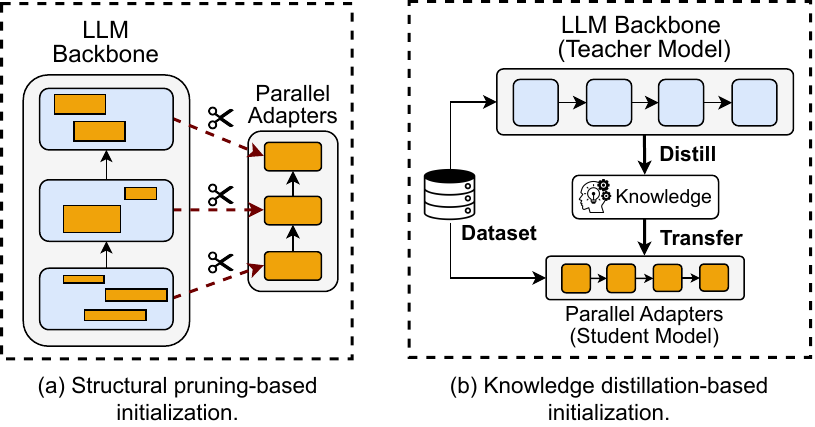}
    \caption{Illustration of weight initialization for Parallel Adapters.}
    \label{fig:init}
    % \vspace{-15pt}
\end{figure}

\subsection{Analysis of Weight Initialization for Parallel Adapters}
\textbf{Observation and Insights.} 
As outlined in \S \ref{sec:paral-adapter}, our Parallel Adapters fine-tune a lightweight proxy network alongside the original LLM backbone. Proper initialization of the proxy network’s weights is essential for accelerating convergence and reducing computational cost and latency, particularly in resource-constrained edge environments.

We begin by observing the PEFT technique employed by LoRA \cite{hu2021lora}. LoRA introduces low-rank matrices $A$ and $B$,  representing the weight update of the original model as the product of these two matrices, i.e., $\Delta W = BA$. 
In forward propagation, the model output is expressed as $h = Wx + \Delta W x=W x + BAx$. In backward propagation, the LLM backbone weights $W$ are frozen, while the parameters of the low-rank matrices $A$ and $B$ are updated.
At the beginning of fine-tuning, LoRA initialized the learnable parameter $\Delta W=0$, with matrix $A$ initialized using a random Gaussian distribution with zero mean, and matrix $B$ initialized as a zero matrix.
The rationale behind this initialization strategy is to ensure that the PEFT initial state remains as closely aligned as possible with the pre-trained model, minimizing random perturbations to the model output during the early stages of training \cite{houlsby2019parameter, hu2021lora}. This helps ensure a smooth transition from the pre-trained state to the fine-tuning phase.

We design our initialization scheme based on this insight. 
As shown in Figure \ref{fig:pa-detail}, directly applying commonly used random Gaussian initialization or zero initialization to our lightweight parallel adapters results in a significant discrepancy between the initial outputs of the proxy network and the original pre-trained LLM, causing substantial perturbations during the early stages of model fine-tuning.
Therefore, we aim for the initialization technique we employ to align the initial states of the lightweight proxy network with frozen LLM backbone.
Given that our proxy network is designed as a lightweight version of the backbone LLM, its initialization weights can be derived from model compression techniques, specifically structural pruning and knowledge distillation. In this work, we design and analyze two initialization methods based on above two techniques.

\textbf{Structural Pruning-Based Initialization.} 
Structural pruning refers to the process of simplifying architectural components, such as dense layers, attention heads, or convolutional channels, while preserving their overall functionality \cite{li2022pruning, ma2023llm, fang2023depgraph}. 
It is widely employed to achieve model compression, effectively reducing memory footprint and accelerating inference.
We observe that the concept of structural pruning can also be applied to initialize the parameters of our Parallel Adapters, where the pruning algorithm takes the backbone LLM as input and outputs its lightweight version for initialization.
We implemented the pruning algorithm in our study using Torch-Pruning \cite{fang2023depgraph,TorchPruning}, a popular open-source toolkit for structural pruning atop PyTorch. The fundamental insight of the Torch-Pruning toolkit \cite{TorchPruning} lies in using a practical norm-based criterion to quantify parameter importance and prune low-significance weights.

\textbf{Knowledge Distillation-Based Initialization.}
Knowledge distillation is another widely used model compression technique, where knowledge is transferred from larger, more competent teacher models to smaller student models that are more suitable for resource-efficient deployment \cite{hinton2015distilling, gu2024minillm, beyer2022knowledge}. 
Similarly, in the context of our task, the backbone LLM can be regarded as the larger teacher model, while the Parallel Adapters function as the smaller student model.
Specifically, we first prompt the teacher model to generate labels along with rationales justifying the labels for an open dataset. We then leverage these labels, in addition to the rationales, to train smaller student models \cite{hsieh2023distilling}. 
We implemented our distillation method based on a popular distillation toolkit open-sourced by Google Research \cite{google_research_distillation}.
In practice, since the distillation process does not require the use of private user data, it can be conducted in cloud data centers equipped with more powerful computational resources.

\begin{figure}[t!]
    \setlength{\abovecaptionskip}{0.1cm}
    \centering
    \includegraphics[width=\linewidth]{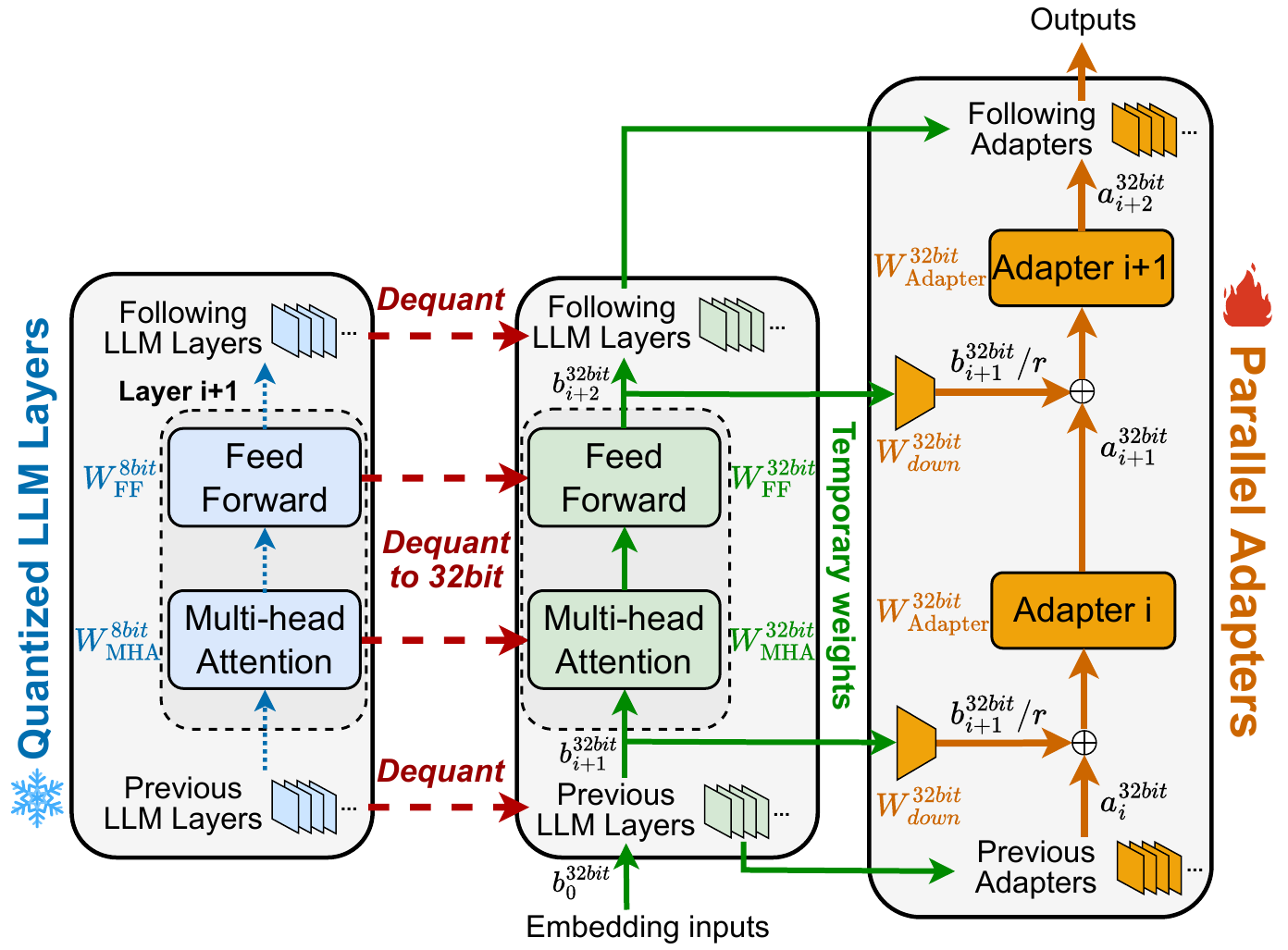}
    \caption{\revision{The workflow of mixed-precision computation performed by our Parallel Adapters.}}
    \label{fig:quan}
    % \vspace{-15pt}
\end{figure}

\subsection{Further Memory Efficiency through Fine-Tuning with Quantized LLMs}
\revision{The memory footprint of LLM fine-tuning is primarily contributed by the \textit{backbone LLM parameters}, \textit{intermediate activations}, and \textit{parameter gradients}.} Our Parallel Adapters design reduces gradient memory overhead by significantly lowering the number of trainable parameters and minimizes intermediate activation memory overhead by avoiding backpropagation through the backbone model.
However, Parallel Adapters do not account for the memory footprint of these substantial backbone model parameters, which can be substantial, especially for LLMs with parameter counts often exceeding hundreds of millions.
Quantization reduces memory usage by mapping model weights from higher (e.g., 32-bit or 16-bit floating points) to lower (e.g., 8-bit or 4-bit integers) bit-width representations. This technique is widely adopted for model compression in LLMs serving and fine-tuning \cite{dettmers2022gpt3, dettmers2024qlora, dettmers2023case, zhang2024quantized}. 
For example, quantizing a 32-bit floating-point (FP32) tensor $\mathbf{X}^{32bit}$ into an 8-bit integer (INT8) tensor $\mathbf{X}^{8bit}$:
\begin{equation}
\begin{aligned}\mathbf{X}^{8bit}=&\mathrm{round}\left(\frac{127}{\mathrm{absmax}(\mathbf{X}^{32bit})}\mathbf{X}^{32bit}\right)\\=&\mathrm{round}\left(c^{32bit}\cdot \mathbf{X}^{32bit}\right),
\end{aligned}
\label{equ:quan}
\end{equation}
where $127$ is the maximum value of the INT8 data type and $\mathrm{absmax}(\mathbf{X}^{32bit})$ represents the maximum absolute value in $\mathbf{X}^{32bit}$. $c^{32bit}$ is the quantization constant for 32-bit tensor $\mathbf{X}^{32bit}$. 
Correspondingly, dequantization is given by: 
\begin{equation}
\mathrm{dequant}(c^{32bit}, \textbf{X}^{8bit})=\frac{\textbf{X}^{8bit}}{c^{32bit}}=\textbf{X}^{32bit}.
\label{equ:dequan}
\end{equation}
This quantization method can be intuitively understood as mapping the range of values representable by FP32 to the range of INT8. 
However, the issue with this approach is that the presence of large-magnitude values (i.e., outliers) in the input tensor leads to inefficient utilization of the available bit combinations, resulting in precision loss. To address the outlier issue, recent pioneering works \cite{dettmers2024qlora} inspire the adoption of block-wise quantization algorithms to reduce quantization loss from large-magnitude outliers in the input tensor. We divide the input tensor $\mathbf{X}$ into contiguous blocks and quantize them independently with Equation \ref{equ:quan}. 

\begin{figure*}[t!]
    \setlength{\abovecaptionskip}{0.1cm}
    \centering
    \includegraphics[width=0.85\linewidth]{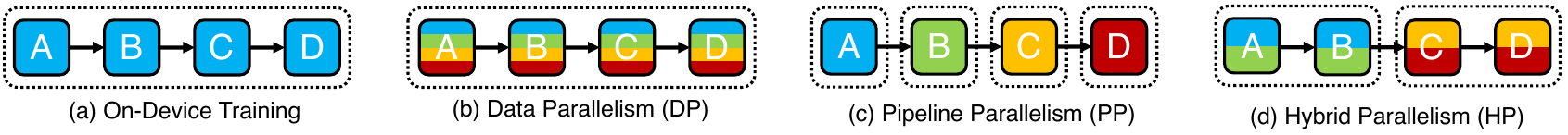}
    \caption{Different parallelism plans of a computation graph, where different colors represent different devices and dashed boxes represent stages. For instance, data parallelism involves ingesting data batches into four devices.}
    \label{fig:parallelism}
    % \vspace{-15pt}
\end{figure*}

In our work, we introduce quantization techniques \cite{dettmers2024qlora} to reduce the significant memory footprint of backbone LLM parameters and perform our fine-tuning algorithm in a mixed-precision computation paradigm, as illustrated in Figure \ref{fig:quan}.
We have one storage data type (typically a lower-bit data type, e.g., INT8) and a computation data type (typically a higher-bit data type, e.g., FP32).
Initially, we quantize the backbone LLM to an 8-bit format for memory-efficient storage, while retaining the lightweight Parallel Adapters in 32-bit precision. During the forward and backward passes, we dequantize the backbone LLM from storage data type to the computation data type to perform FP32-based tensor operations.
\revision{This mixed-precision design simultaneously offers substantial memory reduction (INT8) on the backbone LLM while provisioning high-precision (FP32) tensors for fine-tuning operations.}

\section{Collaborative Edge AI System for Efficient Personal LLMs Fine-Tuning}
In \texttt{PAC+}, we leverage edge devices in physical proximity and associate them as a resource pool to boost in-situ fine-tuning. 
Specifically, the fine-tuning procedure comprises two phases: (1) In the initial epoch, the backbone of LLMs, enhanced with Parallel Adapters, undergoes fine-tuning across multiple edge devices through a blend of data and pipeline parallelism (\S \ref{sec:hybrid-paral}); (2) In subsequent epochs, the activation cache eliminates the necessity for forward propagation within the backbone, thereby enabling the exclusive fine-tuning of our Parallel Adapters utilizing data parallelism (\S \ref{sec:sys-act-cache}).

\subsection{Resource-Efficient Collaborative Orchestration for LLMs Fine-Tuning}
\label{sec:hybrid-paral}
\textbf{Observation of Data and Pipeline Parallelism at the Edge.} 
When collaborating on LLM fine-tuning among edge devices, the principle question is which type of parallelism should be used. 
We analyze different parallelism plans in Figure \ref{fig:parallelism}.
The most common way to train models in parallel is \textit{data parallelism} (DP) \cite{hao2021eddl}. However, DP necessitates that each device maintains a replica of the entire model, a requirement difficult to meet for LLMs with extensive parameter sizes, often surpassing the capacity of a single device. \textit{Pipeline parallelism} (PP) \cite{ye2022eco} is further proposed to address this problem. In PP, the model is partitioned into multiple consecutive stages and each stage is mapped to a separate device. Consequently, PP enables the training of increasingly large models by deploying more devices. 
Nonetheless, PP encounters scalability constraints as the addition of edge devices results in more stages. This not only results in a significant presence of pipeline bubbles but also amplifies the impact of inter-stage communication latency, thereby hindering efficiency. The above observation motivates us to employ a \textit{hybrid parallelism} (HP) architecture that incorporates the best of both DP and PP, so as to achieve superior performance and scalability in resource-constrained edge environments.

\begin{figure}[t]
    \setlength{\abovecaptionskip}{0.1cm}
    \setlength{\belowcaptionskip}{-0.5cm}
    \centering
    \subfigure[The LLM transformer layers is partitioned into two stages, where both Stage 0 and 1 are replicated on a device group with two devices for intra-stage data parallelism.]{
        \begin{minipage}[t]{\linewidth}
        \centering
        \includegraphics[width=0.8\linewidth]{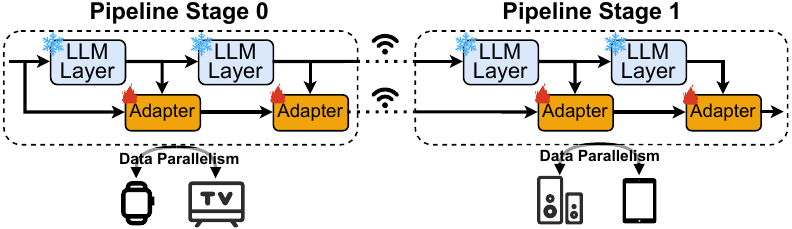}
        \label{fig:stage_partition}
        \end{minipage}
    }
    \subfigure[Fine-tuning pipeline of 6 micro-batches. The numbers in the cells represent micro-batch ids. AllReduce (AR) is performed in both Stage 0 and 1 for model synchronization.
    ]{
        \begin{minipage}[t]{\linewidth}
        \centering
        \includegraphics[width=0.95\linewidth]{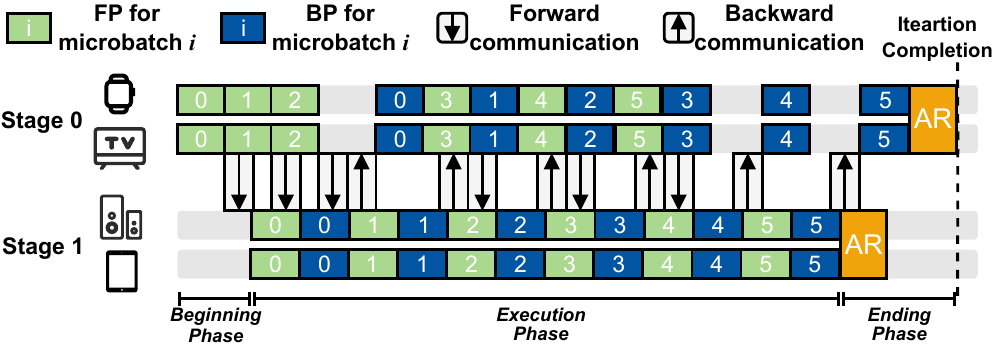}
        \label{fig:hp_pipeline}
        \end{minipage}
    }
    \vspace{-0.2cm}
    \caption{An instance of hybrid parallelism in \texttt{PAC+}.}
    \label{fig:hybrid_parallelism}
    % \vspace{-5pt}
\end{figure}

\textbf{Hybrid Parallelism Architecture in \texttt{PAC+}.}
As illustrated in Figure \ref{fig:stage_partition}, \texttt{PAC+} first divides an LLM into multiple \textit{stages} where each contains a \textit{stage model} composed of a set of consecutive transformer layer. 
Edge devices are allocated into several device groups, each comprising one or more devices. \texttt{PAC+} maps each stage to a group, with the stage model replicated across all devices within that group.
Throughout the fine-tuning process, a mini-batch is divided into several \textit{micro-batches} for concurrent processing to enhance parallelism. If a device cluster hosts multiple devices, micro-batches are further subdivided. Each device is responsible for executing the forward (FP) and backward passes (BP) for its assigned stage model and aggregates gradients across all micro-batches for every mini-batch. Upon completing a mini-batch, gradient synchronization within each device group is achieved through \textit{AllReduce}. Since the majority of parameters in LLMs are frozen, AllReduce synchronizes only the lightweight parallel adapters, ensuring a swift process.
We adopt the \textit{one-forward-one-backward} ($1$F$1$B) micro-batch scheduling \cite{narayanan2019pipedream} which schedules the BP early to release the activation memory produced by FP for reuse. Figure \ref{fig:hp_pipeline} depicts a well-structured hybrid parallelism, encompassing FP, BP, and inter-stage communication.

\textbf{Profiling.} To enable parallelism planning, $\texttt{PAC+}$ profiler first fine-tunes the target LLM using calibration datasets to record the runtime profile required for planning. We define $t_f^{d,l}(\beta)$ and $t_b^{d,l}(\beta)$ as the FP and BP execution times for layer $l$ on device $d$ with batch size of $\beta$, respectively. $u_d$ denotes the memory budget of device $d$.
The size of output activations, input gradients, and weight parameters in bytes will also be collected to calculate memory footprint.

\begin{table}[t!]\setlength{\tabcolsep}{5pt}
\centering
\caption{Table of Notations for Parallelism Planning}
\vspace{-0.3cm}
\label{tab:notation}
\begin{tabularx}{\columnwidth}{cX}
\toprule[1pt]
Notation & \multicolumn{1}{c}{Definition} \\
\hline
$L$ & Total number of layers in the DNN model. \\ 
$\mathcal{D}$ & An ordered set of all devices involved in training. \\ 
$\mathcal{D}_n$ & The subset of first $n$ devices in $\mathcal{D}$. \\ 
$M$ & Number of micro-batches in a mini-batch. \\
$B$ & Size of each micro-batch. \\
$u_d$ & Memory budget available on device $d$. \\ 
$m_d$ & The sum of the memory footprint of LLM parameters, parameter gradients, and activations of device $d$. \\
$G_n$ & A device group involving $n$ edge devices. \\
$t_{f}^{d,l}(\beta)$ / $t_{b}^{d,l}(\beta)$ & Time to perform FP / BP for layer $l$ on device $d$ with a batch size of $\beta$. \\ 
$c_f^s(i)$ / $c_b^s(i)$ & FP / BP communication time between stage $i$ and $i+1$. \\
$W_s$ & Balanced paratition configurations for pipeline with $s$ stages. \\
$e_f^s(i)$ / $e_b^s(i)$ & Time taken for FP / BP in step $i$ with $W_s$. \\
$L_b^s$ / $L_e^s$ / $L_n^s$ & Execution time for the Beginning / Execution / Ending phases in $W_s$. \\
$\text{AR}^s(i)$ & AllReduce time of stage $i$ in $W_s$. \\
\toprule[1pt]
\end{tabularx}
% \vspace{-0.2cm}
\end{table}

\textbf{Planning Algorithm for Hybrid Parallelism.} 
The global throughput of a pipeline is determined by the execution time of the slowest stage. Consequently, our algorithm endeavors to partition the model into balanced stages.
We consider an LLM consisting of $L$ layers and denote $\mathcal{D}$ as an ordered set of all devices involved in planning, while $\mathcal{D}_n = \{d_0, ...d_{n-1}\}$ as the subset of first $n$ devices in $\mathcal{D}$. $W(x\rightarrow y, \mathcal{D}_n, s)$ denote the time taken by the slowest stage in the optimally balanced sub-pipeline between layer $x$ to $y$ with $\mathcal{D}_n$, when divided into $s$ stages. 
To solve this partitioning problem, we break the pipeline into sub-pipelines and leverage the idea of dynamic programming. The formula of the dynamic programming algorithm can be written as:
\begin{equation}
\begin{aligned}
& W(0 \rightarrow y, \mathcal{D}_n, s)= \min _{0 \leqslant q<y} \min _{1 \leqslant m<n} \max \{ W(0 \rightarrow q, \\ & \mathcal{D}_{n-m}, s-1), T(q+1 \rightarrow y,\{d_{n-m} \ldots, d_{n-1}\})\},
\end{aligned}
\label{equ:dp-main}
\end{equation}
where the first term inside the max is the time of the optimally balanced sub-pipeline between layers $0$ to $q$ with $n-m$ devices. 
The second term represents the time required by the single stage comprising layers $q+1$ to $y$ across $m$ devices. The notation $T(x \rightarrow y, \mathcal{G})$ denotes the time required for a single stage to execute FP and BP in a data-parallel manner across the device group $\mathcal{G}$. 
In allocating samples of a mini-batch to devices within $\mathcal{G}$, the goal is to minimize the inference latency determined by the slowest device while ensuring that individual devices do not encounter out-of-memory (OOM) exceptions.
We define $H_{x\rightarrow y}(b, \mathcal{G}_n)$ as the optimal time required by the slowest device in the device group $\mathcal{G}_n = \{d_0,d_1,...,d_{n-1}\}$ to execute the stage model (layers from $x$ to $y$) when distributing $b$ samples across the group.
To solve the aforementioned sample dispatching problem, we also employ a dynamic programming algorithm to search for the optimal sample dispatching strategy. The transition equation for the dynamic programming algorithm can be written as:
\begin{equation}
H_{x\rightarrow y}\left(b, \mathcal{G}_n\right)=\min_{0\le i\le b}\max \left\{
\begin{array}{ll}
H_{x\rightarrow y}(b-i, \mathcal{G}_n\setminus \{d_{n-1}\}), \\
\displaystyle \sum_{l=x}^y \left [t_f^{d_{n-1}, l}(i)+t_b^{d_{n-1}, l}(i)\right ].
\end{array}
\right.
\label{equ:dp2}
\end{equation}

By solving Equation \ref{equ:dp2} with dynamic programming, we ultimately obtain $H_{x \rightarrow y}(B, \mathcal{G})$, which represents the desired value of $T(x \rightarrow y, \mathcal{G})$, where $B$ denotes the micro-batch size.
We define the peak memory footprint of device $d$, denoted as $m_d$, as the sum of memory used by the LLM parameters, parameter gradients, and intermediate activations. 
Increasing the number of samples allocated to a device enlarges the intermediate activations, potentially leading to OOM exceptions.
To exclude OOM cases during the allocation algorithm, the FP and BP time for device $d$ is set to positive infinity. If the algorithm ultimately yields a result of positive infinity for $H_{x \rightarrow y}(B, \mathcal{G})$, it indicates that the collective memory of this group of edge devices cannot accommodate the target stage model.
During the dynamic programming, we will record pipeline planning configurations, including \textit{LLM segmentation points, micro-batch splitting results}, and \textit{device groupings strategies}.

\begin{algorithm}[t]
\small
\setlength{\textfloatsep}{0.5cm}
\setlength{\floatsep}{0.5cm}
\setlength{\intextsep}{-1em} 
\caption{Planning for Hybrid Parallelism}\label{alg:dynamic_programming}
\KwIn{The target LLM model with $L$ layers. An ordered set of all devices $\mathcal{D}$.}
\KwOut{The optimal planning configurations $W_{\sigma}$.}
\SetKwRepeat{Do}{do}{while}
\SetKw{Return}{Return}
\SetKwProg{Ft}{Function}{:}{}

\For{$s \gets 1$ \KwTo $min(L,|\mathcal{D}|)$}{
    \For{$y \gets 1$ \KwTo $L$}{
        \For{$n \gets 1$ \KwTo $|\mathcal{D}|$}{
            \For{$q \gets 0$ \KwTo $y$}{
                \For{$m \gets 0$ \KwTo $n$}{
                    Get $T(q+1\rightarrow y, \{d_{n-m},...,d_{n-1}\})$ by solving Eq. (\ref{equ:dp2})\;
                }
            }
            Update $W(0\rightarrow y, \mathcal{D}_n, s)$ with Eq. (\ref{equ:dp-main})\;
        }
    }
    Calculate $L_b^s$, $L_e^s$, $L_n^s$ using Eq. (\ref{equ:Lb}) and (\ref{equ:Ln})\;
    Update the optimal stage number $\sigma$ with Eq. (\ref{equ:sigma})\;
}
\textbf{Return} $W_{\sigma}$\;
% \vspace*{-0.5em}
\end{algorithm}

Upon the completion of dynamic programming process, we obtain a set of balanced partition configurations for various number of pipeline stages: $\{W_s | \text{ config. of } W(0\rightarrow L, \mathcal{D}, s), s\in\{1,2,...,|\mathcal{D}|\}\}$. The next step is to determine the optimal number of stages. 
Using recorded configurations, we can profile FP and BP execution time of stage $i$ in $W_s$ as $e_f^s(i)$ and $e_b^s(i)$. Similarly, forward and backward communication time between stages $i$ and $i+1$ are represented as $c_f^s(i)$ and $c_b^s(i)$. $\text{AR}^s(i)$ represents the AllReduce time of stage $i$ in $W_s$. $M$ is the number of micro-batch. As shown in Figure \ref{fig:hp_pipeline}, we can divide per mini-batch training of $W_s$ into three phases: \textit{beginning phase}, \textit{execution phase}, and \textit{ending phase} with corresponding latency denoted as $L_b^s, L_e^s, L_n^s$:
\begin{align}
\vspace{-15pt}
    L_b^s = \sum_{i=1}^{s-1}[e_f^s(i) + c_f^s(i)], \quad L_e^s = M \cdot (e_f^s(s) + e_b^s(s)),
    \label{equ:Lb}
\end{align}
\vspace{-15pt}
\begin{align}
    L_n^s = \max_{i\in \{1...,s\}}(\text{AR}^s(i) + \displaystyle\sum_{j=i}^{s-1}(e_b^s(j)+c_b^s(j)),
    \label{equ:Ln}
\end{align}
\vspace{-15pt}
\begin{align}
    \sigma = \operatorname*{argmin}_{s} (L_b^s + L_e^s + L_n^s).
     \label{equ:sigma}
\end{align}
Our algorithm aims to minimize this total latency by optimally determining the number of stages $s$.
We summarize our dynamic programming planning in Algorithm \ref{alg:dynamic_programming}.
We remark that our parallelism planning is an offline procedure that runs once before deployment. The time complexity for our dynamic programming algorithm exhibits an upper bound of $O(BL^2|\mathcal{D}|^3)$. In our experiment, the whole planning time is within several minutes on an edge device.

\subsection{Cache-Enabled Collaborative Edge Fine-Tuning of Parallel Adapters}
\label{sec:sys-act-cache}
\textbf{Data-Parallel Fine-Tuning for Parallel Adapters.} 
The computationally lightweight nature of the Parallel Adapters precludes the use of pipeline parallelism to fine-tuning with activation cache, as it would result in unoverlapable inter-stage communication latency. Therefore, we employ data parallelism to exclusively fine-tune our Parallel Adapters.
Specifically, after the first training epoch, the activation cache for all samples is already collected. We then perform collective communication to redistribute the Parallel Adapters parameters and locally cached activations across all devices, ensuring each device receives the complete set of adapter parameters and corresponding activations. 
The devices then utilize this shared information to fine-tune the parallel adapters in a data-parallel manner. 
In our experiments, fine-tuning the BART-Large model on the MRPC dataset for three epochs, the redistribution of parameters and activations only contributed to approximately $8\%$ of the total training time. Notably, the overhead of this process can be further amortized over additional training epochs.
An instance of personal LLMs fine-tuning with activation cache is depicted in Figure \ref{fig:cache}.

\begin{figure}[t!]
    \setlength{\abovecaptionskip}{0.1cm}
    \centering
    \includegraphics[width=\linewidth]{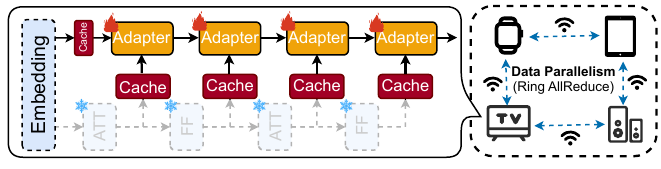}
    \caption{An instance of fine-tuning with activation cache.}
    \label{fig:cache}
    % \vspace{-20pt}
\end{figure}

\textbf{Storage Cost Analysis.} Employing activation caching can reduce the computational requirements of forward propagation; however, it incurs additional storage overhead for activations. Specifically, the storage overhead is $s \times h \times l$ per sequence, where $s$ denotes the sequence length, $h$ represents the transformer's internal feature dimension, and $l$ corresponds to the number of transformer layers. 
For T5-Base model, the activation caching requires less than $1$ GB to store the activations for $500$ training samples with sequence length of $30$. 
Such cost is no more than $1\%$ of the storage of a modern mobile device, e.g., hundreds of GB. During fine-tuning, the activation cache is reloaded from disk per micro-batch, a process that takes no more than tens of milliseconds on embedded flash storage. The cache will be cleared once the fine-tuning process finishes.

\section{Evaluation}
\label{sec:eval}
\subsection{Implementation and Setups}
\textbf{Implementation of \texttt{PAC+}.} We have fully implemented the prototype framework of \texttt{PAC+} and baselines with $\sim$$2,000$ LoC in Python atop Pytorch \cite{pytorch}. \texttt{PAC+}'s idea is also portable and can work well with other lightweight ML frameworks such as MNN \cite{jiang2020mnn} and TF-Lite \cite{tflite}. Our Parallel Adapters is a lightweight version of the backbone model. 
The size of Parallel Adapters is determined by the reduction factor $r$.
All weights and hidden state dimensions of the Parallel Adapters are $\frac{1}{r}$ times the corresponding weights and hidden states of the backbone model.
In our experiments, the reduction factor $r$ is set to 8.	
The weights of the Parallel Adapters are initialized based on structural pruning, using the weights of the backbone model.	
We insert Parallel Adapters at the end of each transformer layer.

\textbf{Models and Datasets.} We evaluate \texttt{PAC+} with three typical transformer based LLM with parameters ranging from $0.25$B to $0.74$B, as detailed in Table \ref{tab:model}, which are widely considered for IPA and edge deployments \cite{li2024personal, yuan2023rethinking}. All experiments were performed under conditions using FP32 precision to ensure fine-tuning performance. We employ two variants of the T5 model \cite{raffel2020exploring}, specifically T5-Base and T5-Large with differing parameter sizes.
We also compare \texttt{PAC+} with baseline methods with BART-Large \cite{lewis2019bart} as the backbone for our parallel adapters.
We evaluate our fine-tuned LLMs with four tasks from GLUE benchmark.
The four tasks evaluate models on multiple diverse tasks over sentiment analysis (SST2), similarity and paraphrase (MRPC, STS-B) and natural language inference (QNLI).

\begin{table}[t!]
\setlength{\tabcolsep}{6.7pt}
% \small
\caption{LLM model specifications used for experiments. "en-de" indicates encoder-decoder LLM structure.}
\vspace{-5pt}
\label{tab:model}
\begin{tabular}{cccccc}
\toprule[1pt]
Model                               & Structure                                                  & Layers & Heads & \begin{tabular}[c]{@{}c@{}}Hidden \\ Size\end{tabular} & \begin{tabular}[c]{@{}c@{}}Param.\\ Count\end{tabular} \\ \hline \hline
T5-Base \cite{raffel2020exploring}  & \begin{tabular}[c]{@{}c@{}}en-de\end{tabular} & 12      & 12     & 768                                                      & 0.25B      \\ \hline
BART-Large \cite{lewis2019bart}           & \begin{tabular}[c]{@{}c@{}}en-de\end{tabular}     & 12     &  16     &1024                                                      & 0.41B                                                  \\ \hline

T5-Large \cite{raffel2020exploring} & \begin{tabular}[c]{@{}c@{}}en-de\end{tabular} & 24     & 16     & 1024                                                      & 0.74B                                             \\ \toprule[1pt]
\end{tabular}
% \vspace{-25pt}
\end{table}

% \usepackage{multirow}
\begin{table}[t!]
\setlength{\tabcolsep}{4pt}
\caption{Specifications of edge devices in Evaluation.}
\vspace{-5pt}
\label{tab:devices}
\begin{tabular}{ccccc}
\toprule[1pt]
Edge Device & GPU Processor & \begin{tabular}[c]{@{}c@{}}Memory\\ Budget\end{tabular} & \begin{tabular}[c]{@{}c@{}}GPU\\ Frequency\end{tabular} & \begin{tabular}[c]{@{}c@{}}Power\\ Mode\end{tabular} \\ \hline \hline
\multirow{2}{*}{Jetson Nano \cite{jetson-nano}} & \multirow{2}{*}{128-core Maxwell} & \multirow{2}{*}{4GB} & 921MHz & 10W (\textbf{H}) \\ \cline{4-5} 
 &  &  & 640MHz & 5W (\textbf{L}) \\ \hline \hline
\multirow{2}{*}{Jetson TX2 \cite{jetson-TX2}} & \multirow{2}{*}{256-core Pascal} & \multirow{2}{*}{8GB} & 1.3GHz & 15W (\textbf{H}) \\ \cline{4-5} 
 &  &  & 850MHz & 7.5W (\textbf{L}) \\ \toprule[1pt]
\end{tabular}
\end{table}

% \usepackage{multirow}
\begin{table*}[t!]\setlength{\tabcolsep}{5.2pt}
\centering
% \small
\caption { Training durations (in hours) for different methods: 3 epochs for MRPC and STS-B, and 1 epoch for SST-2 and QNLI.}
\vspace{-5pt}
\begin{tabular}{cccccccccccccccc}
\toprule[1pt]
\multirow{2}{*}{\begin{tabular}[c]{@{}c@{}}Fine-tuning\\ Techniques\end{tabular}} & \multirow{2}{*}{\begin{tabular}[c]{@{}c@{}}Baseline\\ Methods\end{tabular}} & \multicolumn{4}{c}{T5-Base} &  & \multicolumn{4}{c}{BART-Large} &  & \multicolumn{4}{c}{T5-Large} \\ \cline{3-6} \cline{8-11} \cline{13-16} 
 &  & MRPC & STS-B & SST-2 & QNLI &  & MRPC & STS-B & SST-2  & QNLI  &  & MRPC & STS-B &SST-2 & QNLI \\ \hline \hline

\multirow{3}{*}{Full Model} & Standalone & OOM &OOM  &OOM  & OOM &   & OOM &OOM  & OOM & OOM &  & OOM &OOM  &  OOM& OOM \\
 & PP (Eco-FL) &  0.45 & 0.71 &   2.74& 4.32 &    & 2.41 & 3.78 & 14.56 & 22.98 &  & OOM &  OOM& OOM & OOM \\
 & DP (EDDL) & OOM  & OOM & OOM & OOM &   &  OOM& OOM & OOM & OOM &   & OOM & OOM &  OOM& OOM \\ \hline

\multirow{3}{*}{Adapters} & Standalone & 1.21 & 1.9  & 7.29 & 11.51 &  & OOM & OOM & OOM & OOM &  & OOM & OOM & OOM & OOM \\
 & PP (Eco-FL) & 0.39 & 0.61 & 2.35 & 3.71 &  & 0.54 & 0.85 & 3.27 &  5.16 &  &   2.75&  4.31 & 16.59 &  26.19\\
 & DP (EDDL) & 0.34 & 0.53 & 2.06 & 3.25 &  & OOM & OOM & OOM & OOM &  &  OOM&OOM  & OOM & OOM  \\ \hline

\multirow{3}{*}{LoRA} & Standalone & 1.21 & 1.89 & 7.28 & 11.49  &  & OOM & OOM & OOM & OOM &  & OOM & OOM & OOM & OOM \\
 & PP (Eco-FL) & 0.41 & 0.64 &2.45  & 3.87 &  &   0.55& 0.87 &3.33  & 5.26 &  & 2.73  & 4.28 &  16.48 & 26.02 \\
 & DP (EDDL) & 0.31 & 0.48 & 1.86 & 2.94 &  & OOM & OOM & OOM & OOM &  & OOM & OOM & OOM & OOM  \\ \hline \hline

\textbf{Parallel Adapters} & \textbf{PAC+ (Ours)} & \textbf{0.14} &  \textbf{0.22} & \textbf{1.34} & \textbf{2.12} &  &  \textbf{0.29} &  \textbf{0.45} & \textbf{2.69} & \textbf{4.25} &  &  \textbf{0.69} &   \textbf{1.09} & \textbf{8.88} & \textbf{14.02} \\ \toprule[1pt]

\end{tabular}
\label{tab:training_time}
\vspace{-7pt}
\end{table*}

% \usepackage{multirow}
\begin{table*}[]
\caption{Comparison of final performance between different fine-tuning techniques across four datasets. 
We report the average of F1 score and accuracy for MRPC.
We use Pearson-Spearman Correlation as the metric for STS-B.   
 For  SST-2 and QNLI, we report accuracy. The mean value is the average performance of Full Model, Adapters and LoRA.}
\vspace{-5pt}
 
\label{tab:accuracy}
\begin{tabular}{ccccccccccccccc}
\toprule[1pt]
\multirow{2}{*}{\begin{tabular}[c]{@{}c@{}}Fine-tuning\\ Techniques\end{tabular}} & \multicolumn{4}{c}{T5-Base} &  & \multicolumn{4}{c}{BART-Large} &  & \multicolumn{4}{c}{T5-Large} \\ \cline{2-5} \cline{7-10} \cline{12-15} 
 & MRPC & STS-B & SST-2 & QNLI  &  & MRPC & STS-B &  SST-2  & QNLI &  & MRPC & STS-B & SST-2 &  QNLI \\ \hline \hline
Full Model & 89.71 & 90.94 & 94.03  & 93.08 &  & 88.16 & 91.10 & 95.64 & 94.40&  & 92.78  & 91.08 & 95.30 & 93.30 \\
Adapters  &   88.73    &  90.51 &  93.58  &  93.04&  & 86.63 & 90.24 & 94.93 & 93.27 &  & 91.86 &  90.58& 96.10 & 94.07 \\
 LoRA& 86.27 &  90.73    &93.69   & 93.30 &  & 87.46 & 90.36 & 95.23 &  94.48&  & 90.27 & 92.08 & 95.53 & 94.18 \\ 
 Mean Value &   88.24 & 90.73 & 93.77 & 93.14 &  & 87.42 & 90.57 & 95.27 & 94.05 &  & 91.64 & 91.25 & 95.64 & 93.85   \\ \hline \hline
Parallel Adapters (Ours) &     88.24 &  90.43 &  93.46 &  93.25 &  &  87.71 &  90.54 &  95.25 &  93.68 &  &  91.70 &  91.57 &  95.76 &  93.70    \\  
Difference from Mean   & +0.00 & -0.30 & -0.31 & +0.11 &  & +0.29 & -0.03 & -0.02 & -0.37 &  & +0.06 & +0.32 & +0.12 & -0.15 
\\ \toprule[1pt]
\end{tabular}
\vspace{-7pt}
\end{table*}

\textbf{Edge Environment Setup.} 
We conduct experiments using two off-the-shelf edge devices listed in Table \ref{tab:devices}, simulating four heterogeneous platforms by adjusting their power modes.
We evaluate \texttt{PAC+}’s performance in two realistic edge environments, incorporating both homogeneous and heterogeneous configurations. \textit{Homogeneous Environment A} (\textbf{Env. A}) consists of $4\times$Nano-H, while \textit{Heterogeneous Environment B} (\textbf{Env. B}) comprises $1\times$Nano-H, $1\times$Nano-L, $1\times$TX-H and $1\times$TX-L.
We simulate common network conditions in edge environments (e.g., smart homes) by setting the intra-cluster network bandwidth to $1000$Mbps.

\textbf{Baseline Methods.} 
To evaluate the high performance and robustness of our \texttt{PAC+} system, we comprehensively compare it with a wide range of baseline methods, including those we build ourselves and existing end-to-end collaborative edge training systems.

To thoroughly evaluate the advantages of our \texttt{PAC+} in both algorithmic and system design, we select \textbf{3} classic collaborative edge computing paradigms and \textbf{3} widely adopted LLM fine-tuning methods, combining them in pairs to form \textbf{9} distinct baselines. 
The three collaborative edge computing paradigms are as follows:
\begin{enumerate}[leftmargin=*]
    \item \textbf{Standalone} means fine-tuning LLMs on a single edge device. We compare with it to analyze the scalability performance of \texttt{PAC+}.
    \item \textbf{Data Parallelism (DP)} \cite{li2014communication} is a classic cluster computing paradigm that distributes batch data across devices for simultaneous processing. EDDL \cite{hao2021eddl} is a representative work that applies DP to edge computing, and we reference its architecture to implement our baseline.
    \item \textbf{Pipeline Parallelism (PP)} \cite{huang2019gpipe} enables collaborative training across an edge device cluster by segmenting LLMs into sequential stages for pipeline processing. We implement our baseline framework based on the representative work Eco-FL \cite{ye2022eco}.
\end{enumerate}
The three LLM fine-tuning methods are as follows:
\begin{enumerate}[leftmargin=*]
    \item \textbf{Full model fine-tuning} means that all the LLM parameters are updated for a downstream task.
    \item \textbf{LoRA} \cite{hu2021lora} is a widely-used PEFT technique that decomposes the parameter update for a weight matrix into two trainable low-rank matrices. We apply LoRA to $W_q$ and $W_v$ following the settings from \cite{hu2021lora}.  
    \item  \textbf{Adapters} \cite{houlsby2019parameter} is another widely-used PEFT technique that injects small trainable modules at the end of each transformer layer following the settings from \cite{houlsby2019parameter}. 
\end{enumerate}

In addition to our self-built baselines, we also compare \texttt{PAC+} with two existing end-to-end edge training systems optimized for resource heterogeneity in edge environments:
\begin{enumerate}[leftmargin=*]
    \item \textbf{Asteroid} \cite{ye2024asteroid} is a collaborative DNN training framework designed for heterogeneous edge device clusters. It employs a hybrid data and pipeline parallelism architecture, similar to \texttt{PAC+}, but is designed for full-parameter fine-tuning of both vision and language models.
    \item \textbf{HetPipe} \cite{park2020hetpipe} is a distributed training framework designed for heterogeneous consumer-grade GPU clusters. It facilitates asynchronous parallel training by treating sub-groups of GPUs as virtual workers, employing intra-worker PP and inter-worker communication.
\end{enumerate}

\subsection{Analysis of Algorithmic and System Design Benefits}
\textbf{(1) Comparison of End-to-End Elapsed Fine-Tuning Time Between \texttt{PAC+} and Baselines.}
To thoroughly evaluate the strengths of \texttt{PAC+} in both algorithmic and system design, we select three classic collaborative edge computing paradigms and equip them with three prevalent LLM fine-tuning techniques as baselines in the edge environment Env. A. 
\revision{For the Eco-FL, we divided one mini-batch into 4 micro-batches to enhance pipeline concurrency, whereas the Standalone and EDDL methods were fine-tuned strictly at the mini-batch granularity.}
Table \ref{tab:training_time} summarizes the end-to-end fine-tuning elapsed time on the GLUE datasets for both \texttt{PAC+} and the baseline methods.
% Table \ref{tab:training_time} and Table \ref{tab:accuracy} summarize the end-to-end performance comparisons between \texttt{PAC}, the single-device method, and state-of-the-art collaborative edge training methods. 
For fine-tuning smaller datasets in GLUE, MRPC and STS-B, we conduct training over three epochs, with the latter two epochs benefiting from the \texttt{PAC+} activation cache. 
In contrast, for larger datasets in GLUE, STS-2 and QNLI, a single epoch of fine-tuning is sufficient to achieve satisfactory performance.

As shown in Table \ref{tab:training_time}, \texttt{PAC+} significantly speeds up the training process while preserving convergence performance. 
\texttt{PAC+} achieves an acceleration ranging from $1.21\times$ to $5.44\times$ on SST-2 and QNLI without utilizing activation cache.
In comparison to Standalone and DP, these two baselines often encounter Out-of-Memory (OOM) issues, even when integrating PEFT techniques such as LoRA and Adapters.
This issue stems from the training requirement for each edge device to host the entire target model. 
Particularly for T5-Large, a single Jetson Nano is inadequate to accommodate LLM parameters, not to mention the intermediate activations.	
Compared to PP, \texttt{PAC+} achieves an acceleration ranging from $1.21\times$ to $5.41\times$ on SST-2 and QNLI, without utilizing activation cache.
Parallel Adapters not only alleviate the memory footprint of LLM parameters but also intermediate activations.  
Pipeline parallel strategies allow each edge device to host only a portion of the model parameters. However, these devices still bear a substantial memory footprint from intermediate activations, even when employing PEFT technologies such as LoRA and Adapters.
Therefore, the PP approach necessitates the use of smaller micro-batch sizes or a reduction in the number of micro-batches simultaneously input into the pipeline. This results in decreased concurrency in pipeline parallelism and lowers the training throughput.
Moreover, our hybrid parallelism merges the benefits of both data and pipeline parallelism, providing an expanded search space for parallel architectures to accommodate complex edge environments. Our method enables the identification of the most efficient parallel configuration with maximum throughput within the constraints of available resources.
With the integration of our activation cache mechanism, \texttt{PAC+} achieves speedups of up to $8.64\times$ on the MRPC and STS-B datasets. As discussed in \S \ref{sec:activatoin_cache}, our Parallel Adapters constitute a lightweight, independent network.  
We can skip both the forward and backward passes through the LLM backbone, since the required activations have already been calculated and stored.
Consequently, training overhead can be markedly reduced in the second and third fine-tuning epochs.

\textbf{(2) Analysis of Fine-Tuned LLM Performance with \texttt{PAC+}.} 
Table \ref{tab:training_time} demonstrates that our \texttt{PAC+} achieves higher fine-tuning throughput compared to all baselines.
However, it is still necessary to further validate that our \texttt{PAC+} design does not compromise model accuracy in exchange for lower elapsed time.
Table \ref{tab:accuracy} reports the model performance of LLMs fine-tuned using three classic LLM fine-tuning algorithms and our Parallel Adapters approach.
\revision{All accuracy metrics were evaluated on a server environment with sufficient computational and memory resources to ensure consistent performance reporting, irrespective of potential OOM constraints.}
Fine-tuning involves 3 epochs for the smaller MRPC and STS-B datasets, and 1 epoch for the larger SST-2 and QNLI datasets.
We can observe from Table \ref{tab:accuracy} that \texttt{PAC+} achieves comparable or superior performance to full model fine-tuning and baseline PEFT techniques across various models and datasets. 
The largest discrepancy in mean performance metrics between \texttt{PAC+} and these baseline methods is only -0.37, a negligible difference. Notably, \texttt{PAC+} frequently outperforms these methods and achieves the highest performance on the SST-2 dataset with the T5-Large model.

\begin{figure}[t!]
    \centering
    \subfigure[Total time required to fine-tune 1 epoch on the MRPC dataset in Env. B.]{
        \begin{minipage}[t]{\linewidth}
        \centering
        \includegraphics[width=\linewidth]{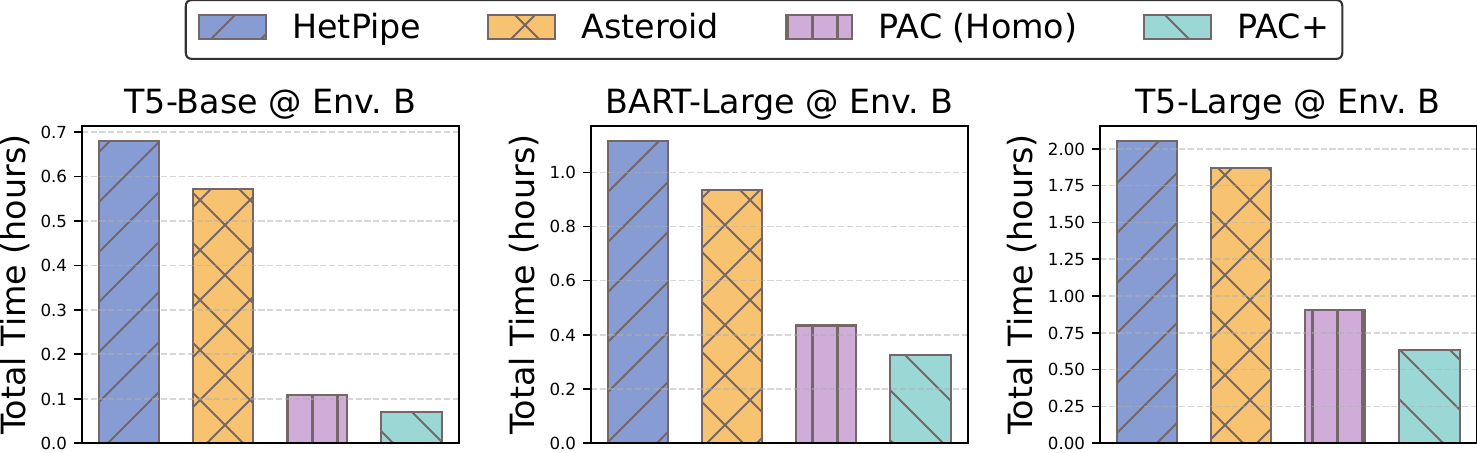}
        \vspace{-0.2cm}
        \label{fig:hete1}
        \end{minipage}
    }
    \subfigure[Total time required to fine-tune 3 epoch on the MRPC dataset in Env. B. \texttt{PAC+} further benefits from the activation cache design.
    ]{
        \begin{minipage}[t]{\linewidth}
        \centering
        \includegraphics[width=\linewidth]{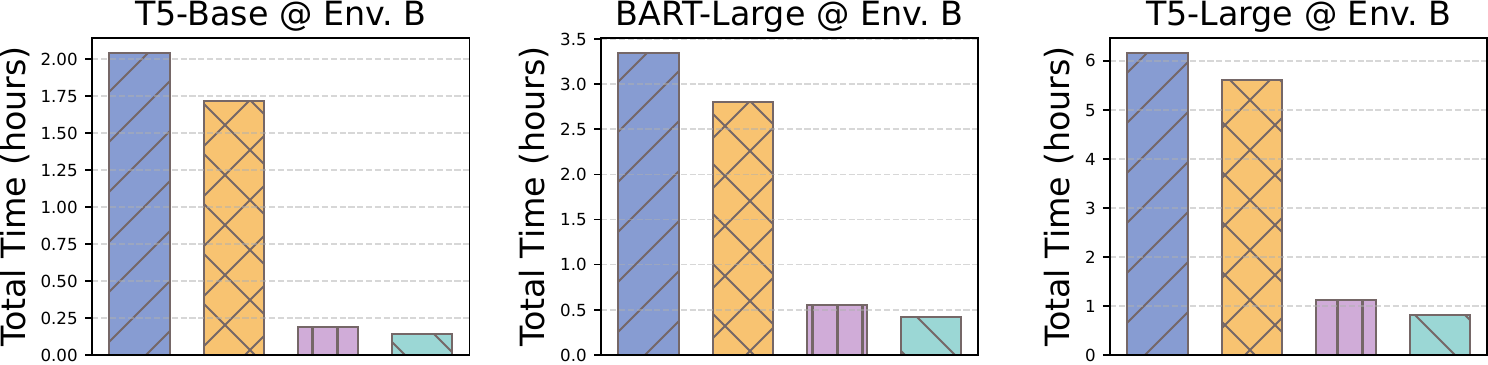}
        \vspace{-0.2cm}
        \label{fig:hete2}
        \end{minipage}
    }
    \vspace{-0.2cm}
    \caption{Comparison of total fine-tuning time across different baselines. \texttt{PAC+} (Homo) refers to a version of the \texttt{PAC+} algorithm that does not consider device resource heterogeneity.}
    \label{fig:significance_Parallel_Adapter}
    % \vspace{-15pt}
\end{figure}

\subsection{Comparison with Existing Collaborative Training Systems under Heterogeneous Edge Environments}
To evaluate the superiority and robustness of our proposed \texttt{PAC+} system compared to existing collaborative edge training systems, we select two state-of-the-art training frameworks designed for heterogeneous device clusters, Asteroid \cite{ye2024asteroid} and HetPipe \cite{park2020hetpipe}, as baselines for comparison with \texttt{PAC+}. The experiments are conducted in the heterogeneous edge environment Env. B.
In addition to comparing \texttt{PAC+} with the two baselines, we also included its older \texttt{PAC} version \cite{ouyang2024pluto}, which does not account for device heterogeneity, as part of our ablation study. This comparison aims to demonstrate the effectiveness of our heterogeneity-aware algorithm design.
We evaluated all three models on the MRPC dataset, conducting fine-tuning for one epoch (Figure \ref{fig:hete1}) and three epochs (Figure \ref{fig:hete2}), respectively.

\begin{figure}[t!]
    \centering
    \subfigure[ The comparison of average sample training time of different fine-tuning techniques. ]{
        \begin{minipage}[t]{\linewidth}
        \centering
        \includegraphics[width=\linewidth]{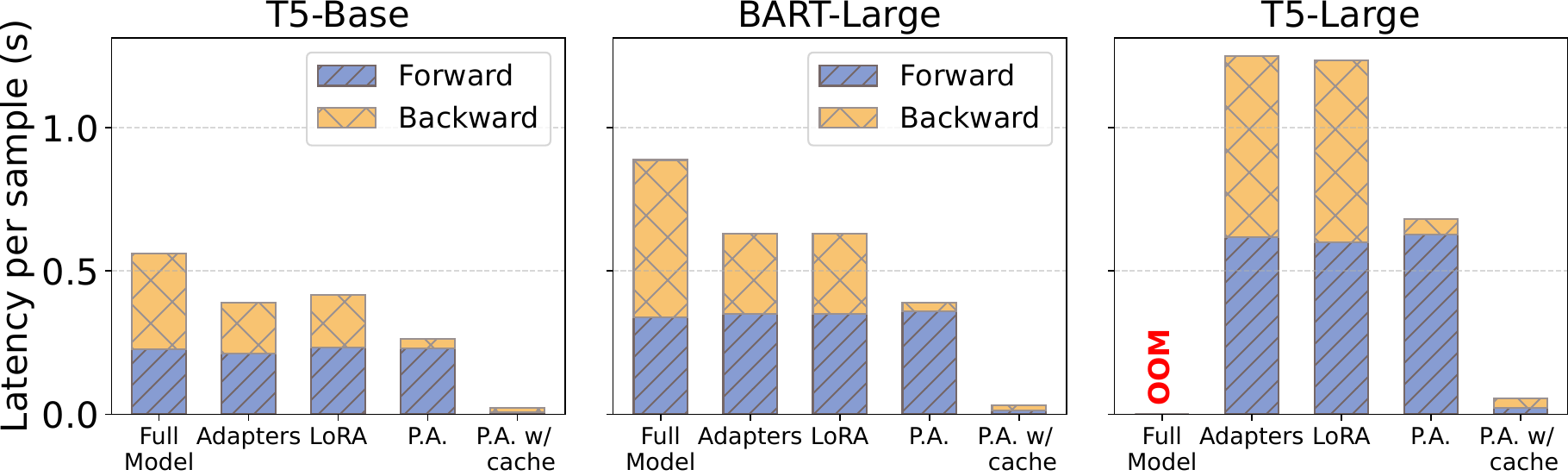}
        \vspace{-0.2cm}
        \label{fig:time_compare}
        \end{minipage}
    }
    \subfigure[Maximum total memory consumption per device across the edge cluster. 
    ]{
        \begin{minipage}[t]{\linewidth}
        \centering
        \includegraphics[width=\linewidth]{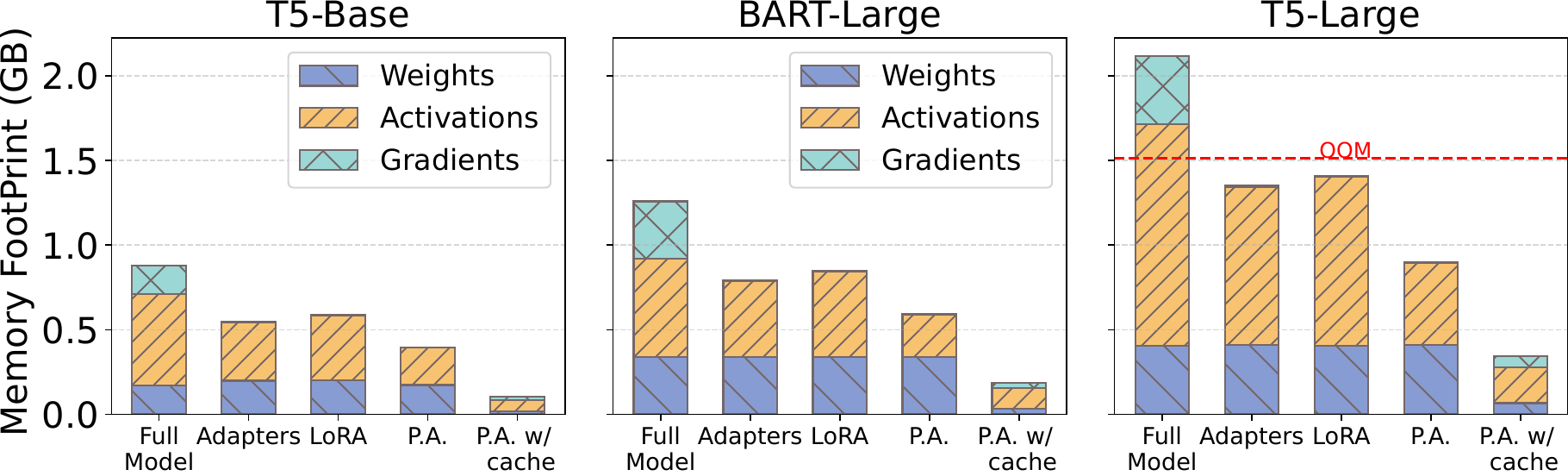}
        \vspace{-0.2cm}
        \label{fig:memory_compare}
        \end{minipage}
    }
    \vspace{-0.2cm}
    \caption{Comparison of different fine-tuning techniques. P.A. indicates our Parallel Adapters technique. Mini-batch size 16; sequence length: 128.}
    \label{fig:significance_Parallel_Adapter}
    % \vspace{-15pt}
\end{figure}

\begin{figure}[t!]
    \centering
    \subfigure[Fine-tuning performance of different initialization strategies on BART-Large.]{
        \begin{minipage}[t]{\linewidth}
        \centering
        \includegraphics[width=\linewidth]{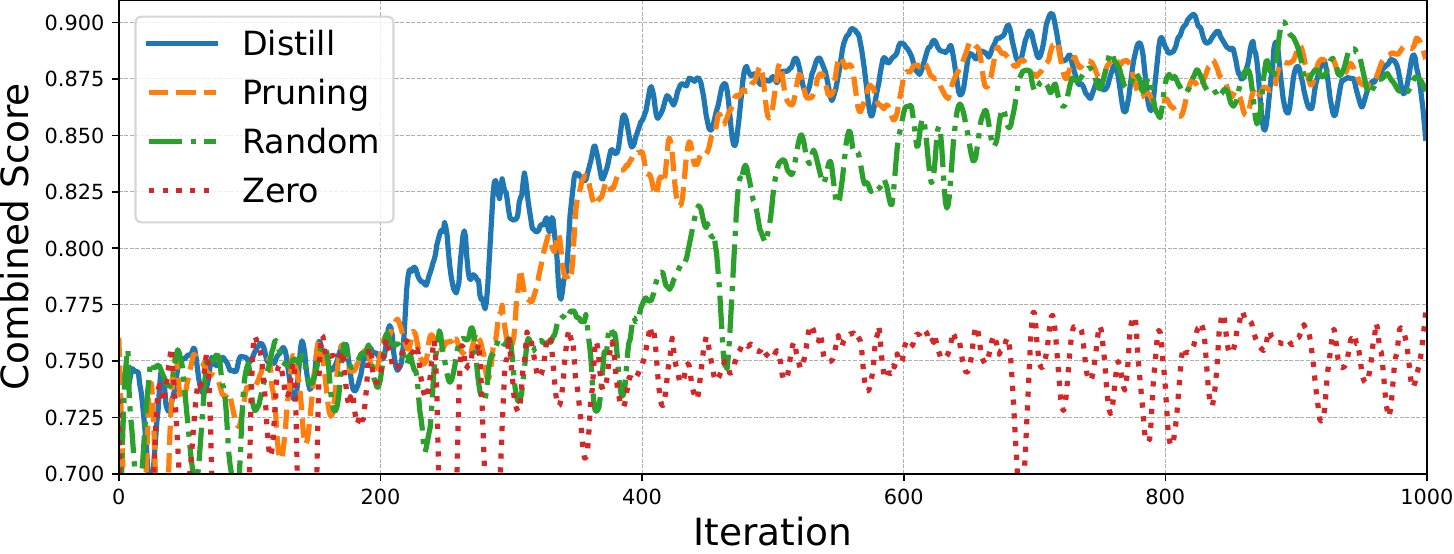}
        \vspace{-0.2cm}
        \label{fig:init_bart}
        \end{minipage}
    }
    \subfigure[Fine-tuning performance of different initialization strategies on T5-Large.
    ]{
        \begin{minipage}[t]{\linewidth}
        \centering
        \includegraphics[width=\linewidth]{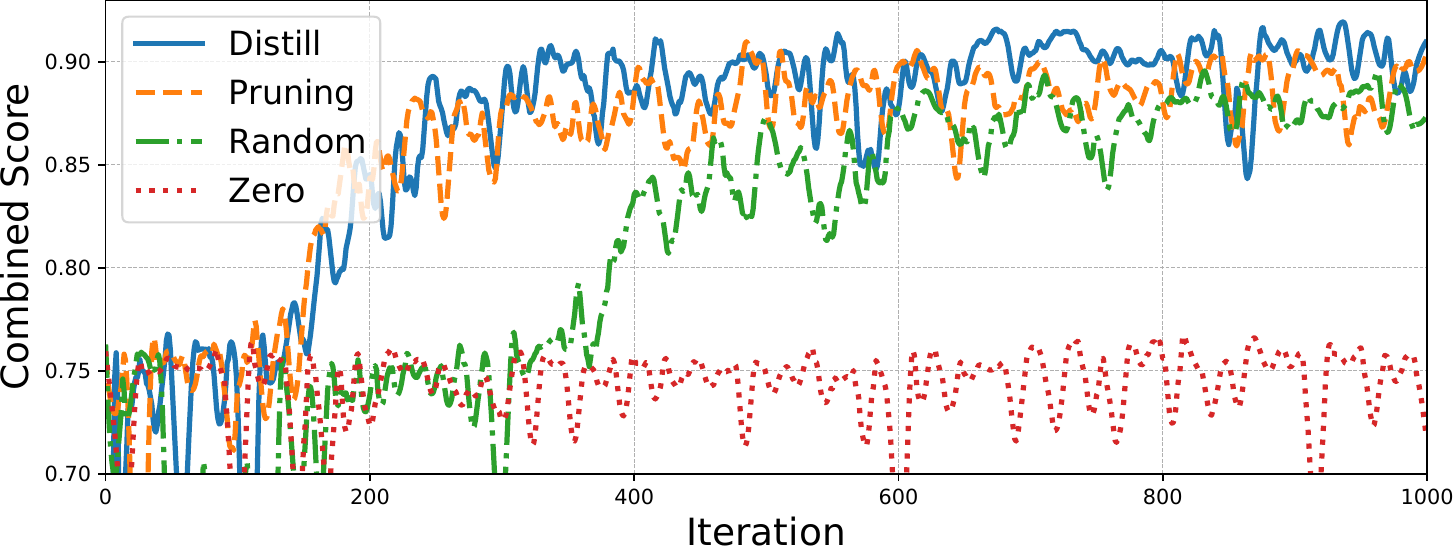}
        \vspace{-0.2cm}
      \label{fig:init_t5}
        \end{minipage}
    }
    \vspace{-0.3cm}
    \caption{The ablation study on different initialization strategies for Parallel Adapters using MRPC dataset.}
    \label{fig:init_PA}
    % \vspace{-10pt}
\end{figure}

\begin{table*}[t!]\setlength{\tabcolsep}{7.1pt}
\caption{Comparison of final performance between different quantization precision across four datasets. We report the average of F1 score and accuracy for MRPC. We use Pearson-Spearman Correlation as the metric for STS-B. For  SST-2 and QNLI, we report accuracy. The mean value is the average performance of Full Model, Adapters and LoRA.}
\vspace{-5pt}
\label{tab:accuracy2}
\begin{tabular}{ccccccccccccccc}
\toprule[1pt]
\multirow{2}{*}{\begin{tabular}[c]{@{}c@{}}Backbone LLM\\ Precision\end{tabular}} & \multicolumn{4}{c}{T5-Base} &  & \multicolumn{4}{c}{BART-Large} &  & \multicolumn{4}{c}{T5-Large} \\ \cline{2-5} \cline{7-10} \cline{12-15} 
 & MRPC & STS-B & SST-2 & QNLI  &  & MRPC & STS-B &  SST-2  & QNLI &  & MRPC & STS-B & SST-2 &  QNLI \\ \hline \hline
FP32 & 88.24 & 90.43 & 93.46  & 93.25 &  & 87.71 & 90.54 & 95.25 & 93.68&  & 91.70  & 91.57 & 95.76 & 93.70 \\
FP16  &   88.42    &  89.33 &  93.35  &  92.58&  & 87.76 & 90.34 & 94.83 & 92.62 &  & 91.24 &  91.17& 95.43 & 92.88 \\
 INT8& 87.45 &  89.12    &92.78   & 91.97 &  & 87.66 & 90.03 & 94.64 &  92.10&  & 91.03 & 90.17 & 94.73 & 92.23 \\ 
 INT4 &   87.24 & 88.25 & 92.66 & 91.10 &  & 87.52 & 89.87 & 94.49 & 91.27 &  & 89.69 & 90.08 & 94.72 & 91.71   \\ 
\toprule[1pt]
\end{tabular}
\vspace{-7pt}
\end{table*}

As shown in Figure \ref{fig:hete1}, under the one-epoch fine-tuning setting, the \texttt{PAC+} architecture achieves a $3.2\times$–$9.7\times$ improvement over HetPipe and a $2.9\times$–$8.1\times$ improvement over Asteroid, attributed to the co-design of the \texttt{PAC+} algorithm and system.
Specifically, HetPipe adopts a hybrid parallelism architecture with inter-group DP and intra-group PP, combining asynchronous updates to address device heterogeneity. However, this parallelism architecture has been shown in existing studies \cite{ye2024asteroid} to require higher inter-device communication, making it unsuitable for bandwidth-constrained edge clusters.
Asteroid adopts the same inter-group PP and intra-group DP architecture as \texttt{PAC+}, along with fine-grained resource partitioning and orchestration to address heterogeneity. However, both Asteroid and HetPipe lacks optimizations for LLM fine-tuning algorithms and relies solely on full-parameter fine-tuning, resulting in higher computational resource requirements and longer training times.
Furthermore, the activation cache design in \texttt{PAC+} enables even greater performance improvements during three-epoch fine-tuning compared to the baselines, as shown in Figure \ref{fig:hete2}. Specifically, \texttt{PAC+} achieves a $7.6\times$–$14.7\times$ improvement over HetPipe and a $6.9\times$–$12.3\times$ improvement over Asteroid.
Our designed \texttt{PAC+} algorithm can achieve up to a $35\%$ latency reduction compared to algorithms that do not account for device heterogeneity.

\subsection{Further Breakdown of the Benefits of Parallel Adapters}
We further conduct experiments to break down the training latency and memory footprint, assessing the time and memory efficiency of Parallel Adapters at the edge.
We conduct experiments on an edge cluster consisting of 8 Jetson Nano-H. All baselines are implemented using the same hybrid parallel architecture and equipped with different fine-tuning algorithms, with 1F1B micro-batch scheduling disabled.
"Activations" contain the intermediate results and optimizer states.
Figure \ref{fig:significance_Parallel_Adapter} illustrates that Parallel Adapters outperform other fine-tuning techniques regarding both time and memory efficiency.

\textbf{(1) Parallel Adapters markedly reduce per-sample training time.}
Figure \ref{fig:time_compare} presents the average sample training time across different fine-tuning techniques. 
Without activation cache, Parallel Adapters can reduce the average sample training time by $31.94\%$ to $56.24\%$ compared to baseline methods, primarily owing to a substantial decrease in backward propagation overhead.
Both Adapters and LoRA incorporate trainable structures into the backbone model, thus necessitating backpropagation across the entire backbone model for gradient computation of these parameters. 
Consequently, regarding backward time, Adapters and LoRA can only achieve approximately a $49\%$ reduction compared to full fine-tuning.
In contrast, backpropagation through the backbone model is unnecessary with Parallel Adapters, leading to a more substantial reduction in backward time, nearly $92\%$ compared to full fine-tuning.
Moreover, Parallel Adapters With activation cache mechanism can further decrease the average sample training time up to $96.39\%$. 
These results demonstrate the substantial reduction in training time achieved by Parallel Adapters.

\begin{figure}[t!]
    \setlength{\abovecaptionskip}{0.1cm}
    \centering
\includegraphics[width=0.8\linewidth]{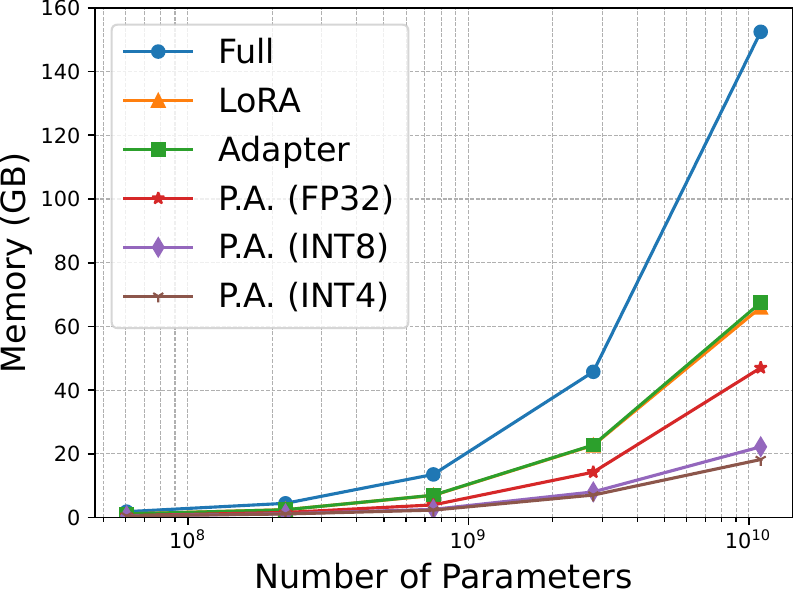}
  \caption{Memory footprint analysis of baselines and Parallel Adapters with various
quantization precision}
    \label{fig:quantize_eval}
    % \vspace{-15pt}
\end{figure}

\textbf{(2) Parallel Adapters yield a substantial reduction in memory usage.} 
Figure \ref{fig:memory_compare} depicts the breakdown of the memory footprint for different fine-tuning techniques.
We report the peak memory consumption per device across edge clusters.
Without activation cache, Parallel Adapters can reduce memory usage by $25.27\%$ to $60.49\%$.
Adapters and LoRA demonstrate parameter efficiency but do not exhibit significant memory efficiency. While these techniques notably decrease the memory footprint of gradients by reducing the number of trainable parameters, the memory usage associated with activations remains considerable.
However, intermediate activations often become the primary memory bottleneck during training, especially with larger batch sizes.
For PEFT techniques such as Adapters and LoRA, activation memory can be reduced by up to $28.15\%$ compared to full fine-tuning across the three models.
In contrast, Parallel Adapters achieve a more significant reduction, reaching up to $58.87\%$.
With activation cache, Parallel Adapters can decrease the peak memory footprint from $74.57\%$ to $88.16\%$ compared to baselines.
This is because it's sufficient to store only the lightweight Parallel Adapters, eliminating the need to host the entire LLM backbone in memory.
% These results demonstrate the substantial reduction in memory usage achieved by Parallel Adapters, which is crucial for practical deployment on mobile devices.

\subsection{Analyzing the Impact of Weight Initialization on Parallel Adapters}
To investigate the impact of weight initialization on the fine-tuning performance of Parallel Adapters, we compare four initialization methods: knowledge distillation, structural pruning, random Gaussian initialization, and zero initialization. The fine-tuning performance of the BART-Large and T5-Large models on the MRPC dataset is illustrated in Figure \ref{fig:init_PA}. 
We implemented our distillation method using Google Research's open-source distillation toolkit \cite{google_research_distillation} and applied structural pruning with Torch-Pruning, a popular open-source toolkit for structural pruning built on PyTorch \cite{fang2023depgraph,TorchPruning}.
Random Gaussian initialization is widely used in popular PEFT techniques, such as LoRA \cite{hu2021lora}.
We can observe from Figure \ref{fig:init_PA} that using distillation or pruning-based initialization for Parallel Adapters allows the model to reach the target accuracy more quickly during fine-tuning compared to random Gaussian or zero initialization. Specifically, using distillation and pruning-based initialization, the BART-Large model achieves a combined score of 0.875 after approximately 410 and 457 iterations, respectively, while random Gaussian initialization requires 605 iterations.
Reducing the required fine-tuning iterations can significantly lower the time and computational resources required for fine-tuning on resource-constrained edge clusters.

\begin{figure}[t]
    \centering
    \subfigure[ Throughput results with a varying number of Jetson Nano.]{
        \begin{minipage}[t]{\linewidth}
        \centering
        \includegraphics[width=\linewidth]{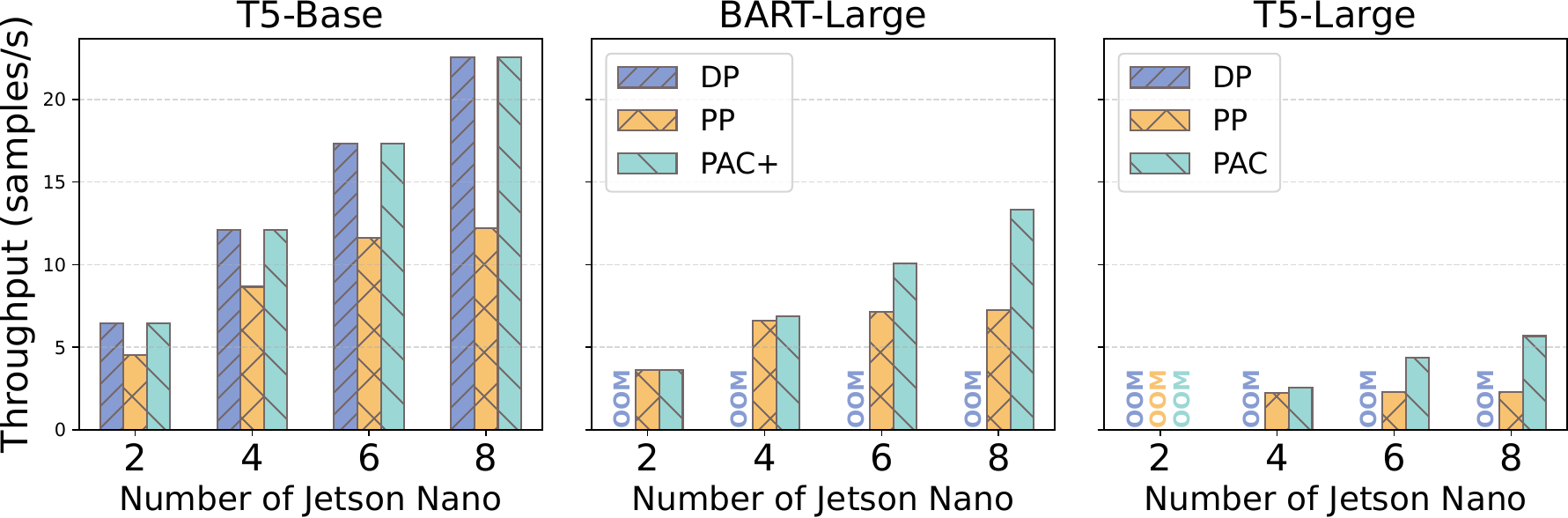}
        \vspace{-0.2cm}
        \label{fig:4.4_throughput}
        \end{minipage}
    }
    \subfigure[Peak memory consumption of LLM weights per device.
    ]{
        \begin{minipage}[t]{\linewidth}
        \centering
        \includegraphics[width=\linewidth]{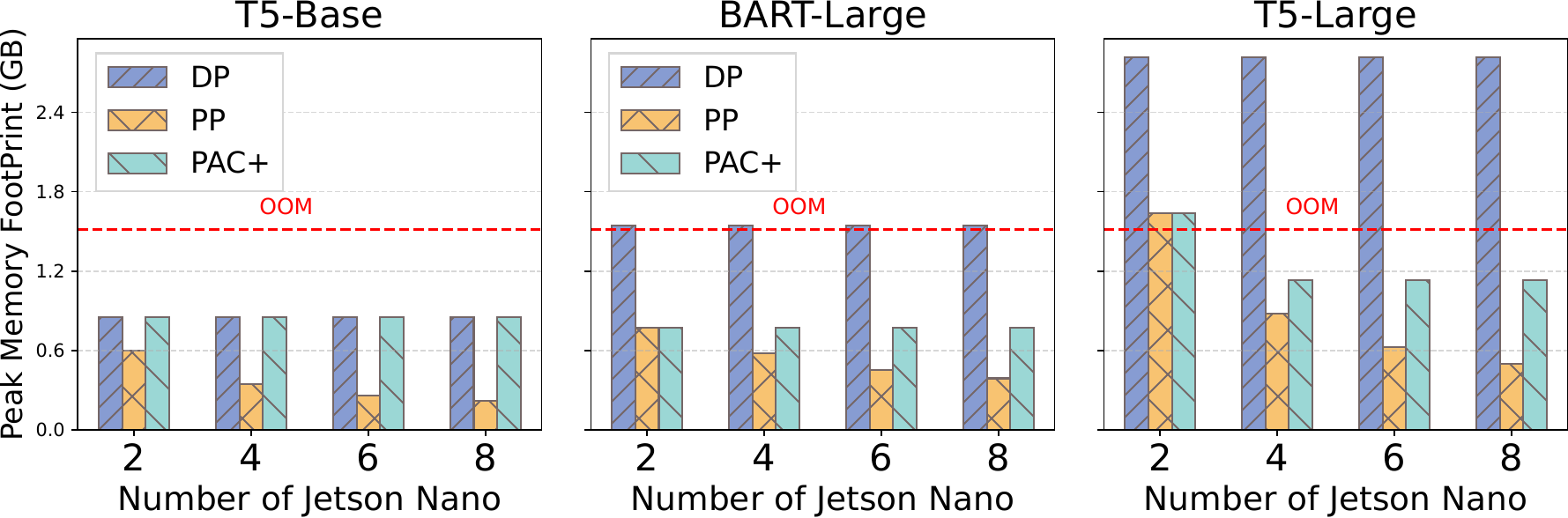}
        \vspace{-0.2cm}
      \label{fig:4.4_memory}
        \end{minipage}
    }
    \vspace{-0.3cm}
    \caption{Comparison of different collaborative edge computing architecture.}
    \label{fig:4.4_pac_eddl_ecofl}
    % \vspace{-10pt}
\end{figure}

\begin{figure}[t!]
    \setlength{\abovecaptionskip}{-2pt}
    \setlength{\belowcaptionskip}{-10pt}
    \centering
    \includegraphics[width=0.95\linewidth]{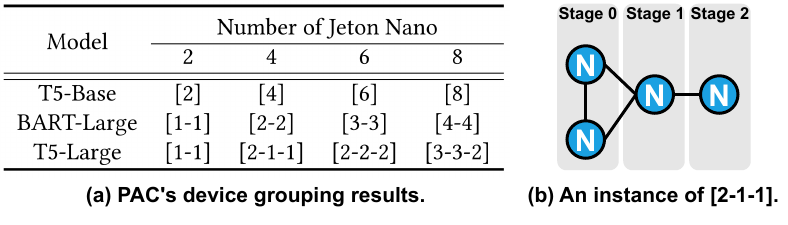}
    \caption {Device grouping results of \texttt{PAC+}'s hybrid parallelism for experiments in Figure \ref{fig:4.4_pac_eddl_ecofl}. "N" indicates Jetson Nano.  }
    \label{fig:4.3_config}
    \vspace{-5pt}
\end{figure}

\subsection{Analysis of Fine-Tuning with Quantized LLMs}
\textbf{(1) Memory Footprint Analysis.}
We compare the memory footprint during the fine-tuning process of LLMs with different number of parameters, using full-parameter fine-tuning, LoRA, Adapter, and Parallel Adapters algorithms with various quantization precisions.
We obtained a series of T5 models with varying parameter sizes by adjusting parameters such as hidden size, number of layers, and number of attention heads.
Experimental results show that Parallel Adapters with lower quantization precisions, such as INT4 and INT8, can significantly reduce the memory footprint during fine-tuning. 
Specifically, compared to full-parameter fine-tuning, the Parallel Adapter algorithm with INT4 precision reduces memory footprint by up to 88\%, and compared to existing PEFT algorithms, it can reduce memory footprint by up to 73\%.

\textbf{(2) Analysis of Fine-Tuning Performance with Quantized LLMs.}
Building on Table \ref{tab:accuracy}, we further evaluate and report the final fine-tuned performance of Parallel Adapters with different quantization precisions across four datasets.
The results are presented in Table \ref{tab:accuracy2}, with the FP32 precision results for Parallel Adapter directly taken from Table \ref{tab:accuracy}.
Experimental results show that fine-tuning with low-precision backbone LLM does not significantly degrade model performance and they still maintain competitive accuracy. Our previous experiments have demonstrated that quantization can significantly reduce memory footprint during fine-tuning, making it suitable for resource-constrained edge environments.

\subsection{Analysis the Benefits and Scalability of the Hybrid Parallelism Architecture.}
Compared to standalone data parallelism or pipeline parallelism, the hybrid parallelism architecture adopted by \texttt{PAC+} integrates the strengths of both DP and PP, enabling superior performance and scalability in resource-constrained edge environments.
To explore the scalability advantages of \texttt{PAC+}'s hybrid parallelism over DP and PP, we compared the throughput of these methods when fine-tuning collaboratively across 2 to 8 Jetson Nano-H devices. The batch size was consistent with the number of devices, and the sequence length of each sample was fixed at 128. We implement DP and PP using the Parallel Adapters technique to ensure a fair comparison. Note that none of the three methods utilizes activation cache.

\begin{figure}[t!]
    \setlength{\abovecaptionskip}{0.1cm}
    \centering
    \includegraphics[width=\linewidth]{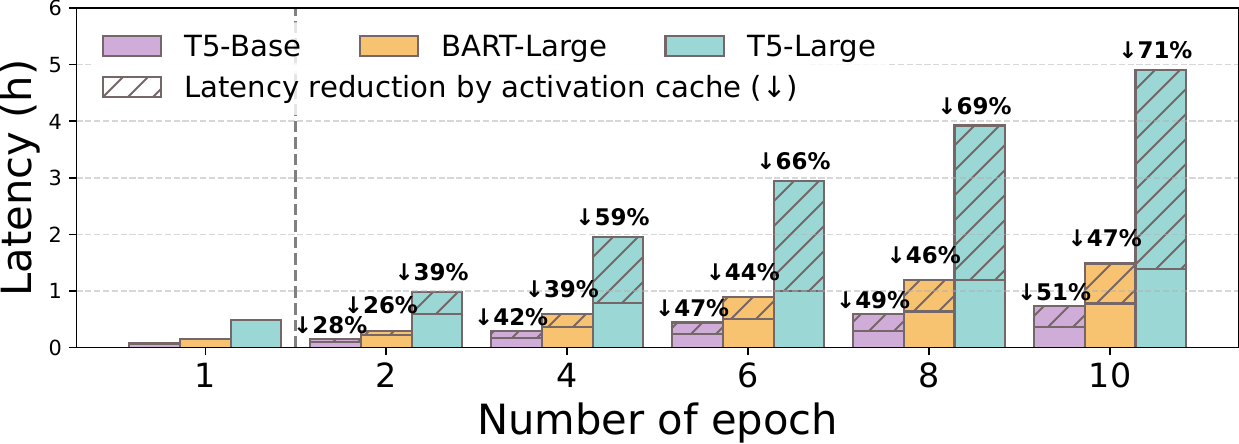}
  \caption{Fine-tuning time with \texttt{PAC+}. Time without activation cache is represented by bars. The corresponding reduction in time achieved utilizing activation cache is represented by shaded areas. Dataset: MRPC.}
    \label{fig:activation_cache}
    % \vspace{-15pt}
\end{figure}

Figure \ref{fig:4.4_memory} illustrates the maximum per device memory footprint of model weights across edge cluster. 
For DP, each device must host a complete LLM, preventing the reduction of the parameters' memory footprint through scaling up the number of devices. 
Therefore, as shown in Figure \ref{fig:4.4_throughput}, the DP method exhibits OOM errors with both the BART-Large and T5-Large models.
Conversely, \texttt{PAC+} and PP partition the model into multiple stages with each handled by different devices. This approach allows for scaling the number of devices to reduce the peak memory footprint. \texttt{PAC+}'s hybrid parallelism offers a broader search space for parallel strategies compared to standalone PP. Our planning algorithm for \texttt{PAC+} is capable of identifying more efficient hybrid parallel configurations within memory constraints, enhancing resource utilization. Although \texttt{PAC+} may incur higher memory overhead in some instances, it achieves greater system throughput. Specifically, when compared to PP, \texttt{PAC+} exhibits an increase in throughput from $39.50\%$ to $84.79\%$.

To more clearly illustrate the parallel strategies adopted by \texttt{PAC+}, we present the device grouping configurations for \texttt{PAC+} across various LLMs and numbers of devices in Figure \ref{fig:4.3_config}. On the left, a table displays all the grouping results across three models. On the right, an instance is shown where a model is divided into three stages, with two devices handling the first stage to perform data parallelism.
Specifically, when fine-tuning BART-Large with eight devices, DP encounters OOM issues because a single Jetson Nano cannot accommodate a complete BART-Large. PP addresses this problem by dividing the model into eight stages and employing straight pipeline parallelism for training.
On the contrary, our \texttt{PAC+} approach divides BART-Large into two stage models, with each stage replicated across four devices. This configuration significantly reduces the number of stages in the pipeline, thereby minimizing inter-stage data dependencies and communication latency, which in turn enhances the pipeline's concurrent efficiency.
These results demonstrate that our hybrid parallel approach offers a larger search space for parallel configurations, providing enhanced scalability and robustness across varying numbers of devices and workloads.

\subsection{Comparison of \texttt{PAC+} with and without Activation Cache.}
We further investigated how our activation cache mechanism benefits the required fine-tuning latency.
By leveraging activation cache, the fine-tuning latency per epoch can decrease up to $79.51\%$. Figure \ref{fig:activation_cache} shows the fine-tuning latency reduction as the number of epochs increases with the use of the activation cache mechanism. We can observe that by leveraging the activation cache, fine-tuning latency is reduced by $26\%$ to $71\%$.
Moreover, this reduction in latency increases with the number of epochs. For example, with the T5-Large model, training for two epochs reduces latency by $39\%$, whereas training for ten epochs increases the reduction to $71\%$.
This reduction can be attributed to the fact that the Parallel Adapters constitute a lightweight, independent network, resulting in a significant decrease in training cost compared to the LLM backbone.
We can bypass both the forward and backward passes through the LLM backbone since the necessary activations are already cached.

\section{Related Work}
\label{sec:related-work}
\noindent\textbf{Parameter-Efficient Fine-Tuning for LLM.} 
% PEFT can achieve similar results to full model fine-tuning while modifying around $1\%$ of the model's parameters. 
Prompt tuning \cite{lester2021power} proposes to prepend the model input embeddings with a trainable tensor. 
Adapters tuning \cite{houlsby2019parameter} adds domain-specific layers after attention and FFN layers in transformer. LoRA \cite{hu2021lora} decomposes the parameter update for a weight matrix into two trainable low-rank matrices. To further reduce the memory overhead, pioneering studies explore fine-tuning techniques that obviate the need for backpropagation through the backbone model. 
Y-tuning \cite{liu2024tuning} learns additional task-specific label representations, which are integrated with the output of the backbone model to circumvent backpropagation. 
LST \cite{sung2022lst} involves the use of pruned lightweight transformer structures from the backbone as a side network. E$^3$VA \cite{yin2023parameter} extends the concept of the side network into the realm of computer vision.

\revision{\textbf{Model Compressing Techniques for LLM Fine-tuning.}
Recent pioneering techniques focus on compressing LLMs to mitigate resource challenges. For instance, structural pruning, exemplified by LLM-Pruner \cite{ma2023llm}, selectively removes non-critical coupled structures based on gradient information, thereby maximally preserving the majority of the LLM’s functionality. Similarly, researchers at Google \cite{google_research_distillation} have demonstrated transferring knowledge from a large billion-parameter T5 teacher model to a student model with only millions of parameters via knowledge distillation, while retaining near-comparable performance. By reducing the model's volume through compression, computational and memory resources required for full parameter fine-tuning can also be decreased.

The Parallel Adapter approach utilized in our work is categorized as a PEFT technique. However, our design is orthogonal to aforementioned model compression methods.
Specifically, when the resource demands for the LLM backbone are substantial, the backbone can initially undergo appropriate compression to fit within the constraints of the target edge environment. Then, our Parallel Adapter technology can be further applied to significantly reduce the resources needed for fine-tuning, while simultaneously providing superior pluggable characteristics and mitigating catastrophic forgetting—critical advantages over full fine-tuning.}

\textbf{On-device DNN Fine-Tuning.} 
POET \cite{patil2022poet} achieves the finetuning of a BERT model on embedded devices, optimizing for both training speed and energy consumption. Lin et al. \cite{lin2022device} enable training directly on devices with a minimal memory requirement of only 256KB. Sage and Melon \cite{gim2022memory, wang2022melon} implement hybrid memory management and conservation strategies, including operator fusion and the use of a dedicated memory pool, to mitigate memory limitations. 
Additionally, Mandheling \cite{xu2022mandheling} incorporates mixed-precision training along with DSP offloading to enhance the speed of learning. 
% Octo \cite{zhou2021octo} introduces a novel INT8 training method incorporating Loss-aware Compensation to boost on-device learning efficiency.

\textbf{Collaborative Edge Computing for DNN Fine-Tuning.} 
Federated Learning (FL) has been a promising paradigm in distributed machine learning that enables in-situ model fine-tuning. FwdLLM \cite{xu2024fwdllm} designs a backpropagation-free fine-tuning FL protocol to enhance efficiency. AdaFL \cite{cai2023efficient} proposes an FL framework for fine-tuning LLMs that features adaptable depth and width in its adapters modules.
Breaking through the conventional paradigm of FL, Ye 
et al. \cite{ye2022eco, zeng2024implementation} devise a pipeline parallel architecture that facilitates the collaborative fine-tuning of DNNs across multiple edge devices. EDDL \cite{hao2021eddl} adopts data parallelism training across embedded devices in a local area network. Asteroid \cite{ye2024asteroid} also employs HPP across multiple edge devices for DNN training, but it does not specifically address the parameter-efficient fine-tuning of LLMs.

This paper is an extension of the shorter conference version \cite{ouyang2024pluto}.
In this work, we further explore low-precision quantization of the LLM backbone and weight initialization for Parallel Adapters, enabling further optimization of both memory and time overhead in LLM fine-tuning. Also, we design a novel heterogeneity-aware parallelism planning algorithm to fully exploit the computational resource of heterogeneous edge clusters. Additionally, we provide a more detailed discussion of the design and implementation of our Parallel Adapters architecture.

% \vspace{-5pt}
\section{Conclusion}
\label{sec:conclusion}
This paper proposes \texttt{PAC+}, a time and memory efficient collaborative edge AI framework for personal LLMs fine-tuning. \texttt{PAC+} breaks the resource wall of personal LLMs fine-tuning with a sophisticated algorithm-system co-design.
Extensive evaluation of the prototype implementation demonstrates that \texttt{PAC+} outperforms existing collaborative edge training systems, achieving up to a $9.7\times$ end-to-end speedup and reducing memory footprint by up to $88.16\%$ compared to mainstream LLM fine-tuning algorithms.

\bibliographystyle{IEEEtran}
\bibliography{reference}

@article{ma2023llm,
  title={Llm-pruner: On the structural pruning of large language models},
  author={Ma, Xinyin and Fang, Gongfan and Wang, Xinchao},
  journal={Advances in neural information processing systems},
  volume={36},
  pages={21702--21720},
  year={2023}
}

@article{dettmers2024qlora,
  title={Qlora: Efficient finetuning of quantized llms},
  author={Dettmers, Tim and Pagnoni, Artidoro and Holtzman, Ari and Zettlemoyer, Luke},
  journal={Advances in Neural Information Processing Systems},
  volume={36},
  year={2024}
}

@article{dettmers2022gpt3,
  title={Gpt3. int8 (): 8-bit matrix multiplication for transformers at scale},
  author={Dettmers, Tim and Lewis, Mike and Belkada, Younes and Zettlemoyer, Luke},
  journal={Advances in Neural Information Processing Systems},
  volume={35},
  pages={30318--30332},
  year={2022}
}

@misc{TorchPruning,
  author = {VainF},
  title = {Torch-Pruning: A Structured Pruning Toolkit for PyTorch},
  year = {2023},
  howpublished = {GitHub repository},
  note = {Available at https://github.com/VainF/Torch-Pruning}
}

@inproceedings{li2022pruning,
  author       = {Hao Li and
                  Asim Kadav and
                  Igor Durdanovic and
                  Hanan Samet and
                  Hans Peter Graf},
  title        = {Pruning Filters for Efficient ConvNets},
  booktitle    = {5th International Conference on Learning Representations, {ICLR} 2017,
                  Toulon, France, April 24-26, 2017, Conference Track Proceedings},
  year         = {2017},
  pages = {1683--1695},
}

@inproceedings{google_research_distillation,
  title={Distilling step-by-step! outperforming larger language models with less training data and smaller model sizes},
  author={Hsieh, Cheng-Yu and Li, Chun-Liang and Yeh, Chih-Kuan and Nakhost, Hootan and Fujii, Yasuhisa and Ratner, Alex and Krishna, Ranjay and Lee, Chen-Yu and Pfister, Tomas},
  booktitle={Findings of the Association for Computational Linguistics: ACL 2023},
  pages={8003--8017},
  year={2023}
}

@inproceedings{beyer2022knowledge,
  title={Knowledge distillation: A good teacher is patient and consistent},
  author={Beyer, Lucas and Zhai, Xiaohua and Royer, Am{\'e}lie and Markeeva, Larisa and Anil, Rohan and Kolesnikov, Alexander},
  booktitle={Proceedings of the IEEE/CVF conference on computer vision and pattern recognition},
  pages={10925--10934},
  year={2022}
}

@inproceedings{gu2024minillm,
  title={MiniLLM: Knowledge distillation of large language models},
  author={Gu, Yuxian and Dong, Li and Wei, Furu and Huang, Minlie},
  booktitle={The Twelfth International Conference on Learning Representations},
  pages={13278--13301},
  year={2024}
}

@article{hinton2015distilling,
  title={Distilling the Knowledge in a Neural Network},
  author={Hinton, Geoffrey},
  journal={arXiv preprint arXiv:1503.02531},
  year={2015}
}

@article{hsieh2023distilling,
  title={Distilling step-by-step! outperforming larger language models with less training data and smaller model sizes},
  author={Hsieh, Cheng-Yu and Li, Chun-Liang and Yeh, Chih-Kuan and Nakhost, Hootan and Fujii, Yasuhisa and Ratner, Alexander and Krishna, Ranjay and Lee, Chen-Yu and Pfister, Tomas},
  journal={arXiv preprint arXiv:2305.02301},
  year={2023}
}

@inproceedings{fang2023depgraph,
  title={Depgraph: Towards any structural pruning},
  author={Fang, Gongfan and Ma, Xinyin and Song, Mingli and Mi, Michael Bi and Wang, Xinchao},
  booktitle={Proceedings of the IEEE/CVF conference on computer vision and pattern recognition},
  pages={16091--16101},
  year={2023}
}

@article{zeng2024implementation,
  title={Implementation of Big AI Models for Wireless Networks with Collaborative Edge Computing},
  author={Zeng, Liekang and Ye, Shengyuan and Chen, Xu and Yang, Yang},
  journal={IEEE Wireless Communications},
  volume={31},
  number={3},
  pages={50--58},
  year={2024},
  publisher={IEEE}
}

@inproceedings{yuan2023rethinking,
  title={Mobile foundation model as firmware},
  author={Yuan, Jinliang and Yang, Chen and Cai, Dongqi and Wang, Shihe and Yuan, Xin and Zhang, Zeling and Li, Xiang and Zhang, Dingge and Mei, Hanzi and Jia, Xianqing and others},
  booktitle={Proceedings of the 30th Annual International Conference on Mobile Computing and Networking},
  pages={279--295},
  year={2024}
}

@inproceedings{narayanan2019pipedream,
  title={PipeDream: generalized pipeline parallelism for DNN training},
  author={Narayanan, Deepak and Harlap, Aaron and Phanishayee, Amar and Seshadri, Vivek and Devanur, Nikhil R and Ganger, Gregory R and Gibbons, Phillip B and Zaharia, Matei},
  booktitle={SOSP},
  pages={1--15},
  year={2019}
}

@article{li2014communication,
  title={Communication efficient distributed machine learning with the parameter server},
  author={Li, Mu and Andersen, David G and Smola, Alexander J and Yu, Kai},
  journal={Advances in Neural Information Processing Systems},
  pages= {19--27},
  volume={27},
  year={2014}
}

@inproceedings{park2020hetpipe,
  title={$\{$HetPipe$\}$: Enabling large $\{$DNN$\}$ training on (whimpy) heterogeneous $\{$GPU$\}$ clusters through integration of pipelined model parallelism and data parallelism},
  author={Park, Jay H and Yun, Gyeongchan and Chang, M Yi and Nguyen, Nguyen T and Lee, Seungmin and Choi, Jaesik and Noh, Sam H and Choi, Young-ri},
  booktitle={2020 USENIX Annual Technical Conference (USENIX ATC 20)},
  pages={307--321},
  year={2020}
}

@article{huang2019gpipe,
  title={Gpipe: Efficient training of giant neural networks using pipeline parallelism},
  author={Huang, Yanping and Cheng, Youlong and Bapna, Ankur and Firat, Orhan and Chen, Dehao and Chen, Mia and Lee, HyoukJoong and Ngiam, Jiquan and Le, Quoc V and Wu, Yonghui and others},
  journal={NeurIPS},
  pages={103--112},
  volume={32},
  year={2019}
}

@article{han2024parameter,
  author       = {Zeyu Han and
                  Chao Gao and
                  Jinyang Liu and
                  Jeff Zhang and
                  Sai Qian Zhang},
  title        = {Parameter-Efficient Fine-Tuning for Large Models: {A} Comprehensive
                  Survey},
  journal      = {Trans. Mach. Learn. Res.},
  volume       = {2024},
  year         = {2024},
}

@article{xu2023llmcad,
  title={EdgeLLM: Fast On-Device LLM Inference With Speculative Decoding},
  author={Xu, Daliang and Yin, Wangsong and Zhang, Hao and Jin, Xin and Zhang, Ying and Wei, Shiyun and Xu, Mengwei and Liu, Xuanzhe},
  journal={IEEE Transactions on Mobile Computing},
  volume={24},
  number={4},
  pages={3256--3273},
  year={2025},
  publisher={IEEE}
}

@inproceedings{guo2023sti,
  title={Sti: Turbocharge nlp inference at the edge via elastic pipelining},
  author={Guo, Liwei and Choe, Wonkyo and Lin, Felix Xiaozhu},
  booktitle={ASPLOS, Volume 2},
  pages={791--803},
  year={2023}
}

@article{li2024personal,
  title={Personal llm agents: Insights and survey about the capability, efficiency and security},
  author={Li, Yuanchun and Wen, Hao and Wang, Weijun and Li, Xiangyu and Yuan, Yizhen and Liu, Guohong and Liu, Jiacheng and Xu, Wenxing and Wang, Xiang and Sun, Yi and others},
  journal={arXiv preprint arXiv:2401.05459},
  year={2024}
}

@inproceedings{dettmers2023case,
  title={The case for 4-bit precision: k-bit inference scaling laws},
  author={Dettmers, Tim and Zettlemoyer, Luke},
  booktitle={International Conference on Machine Learning},
  pages={7750--7774},
  year={2023},
  organization={PMLR}
}

@inproceedings{zhang2024quantized,
  title={Quantized side tuning: Fast and memory-efficient tuning of quantized large language models},
  author={Zhang, Zhengxin and Zhao, Dan and Miao, Xupeng and Oliaro, Gabriele and Zhang, Zhihao and Li, Qing and Jiang, Yong and Jia, Zhihao},
  booktitle={Proceedings of the 62nd Annual Meeting of the Association for Computational Linguistics (Volume 1: Long Papers)},
  pages={1--17},
  year={2024}
}

@inproceedings{zhang2020side,
  title={Side-tuning: a baseline for network adaptation via additive side networks},
  author={Zhang, Jeffrey O and Sax, Alexander and Zamir, Amir and Guibas, Leonidas and Malik, Jitendra},
  booktitle={ECCV 2020, Proceedings, Part III 16},
  pages={698--714},
  year={2020},
  organization={Springer}
}

@inproceedings{wei2024communication,
  title={Communication-Efficient Model Parallelism for Distributed In-situ Transformer Inference},
  author={Wei, Yuanxin and Ye, Shengyuan and Jiang, Jiazhi and Chen, Xu and Huang, Dan and Du, Jiangsu and Lu, Yutong},
  booktitle={DATE},
  pages={1--6},
  year={2024},
  organization={IEEE}
}

@article{ye2024galaxy,
  title={Galaxy: A Resource-Efficient Collaborative Edge AI System for In-situ Transformer Inference},
  author={Ye, Shengyuan and Du, Jiangsu and Zeng, Liekang and Ou, Wenzhong and Chu, Xiaowen and Lu, Yutong and Chen, Xu},
  journal={arXiv preprint arXiv:2405.17245},
  year={2024}
}

@article{vaswani2017attention,
  title={Attention is all you need},
  author={Vaswani, Ashish and Shazeer, Noam and Parmar, Niki and Uszkoreit, Jakob and Jones, Llion and Gomez, Aidan N and Kaiser, {\L}ukasz and Polosukhin, Illia},
  journal={NeurIPS},
  volume={30},
  pages={5998--6008},
  year={2017}
}

@inproceedings{lewis2019bart,
  title={BART: Denoising sequence-to-sequence pre-training for natural language generation, translation, and comprehension},
  author={Lewis, Mike and Liu, Yinhan and Goyal, Naman and Ghazvininejad, Marjan and Mohamed, Abdelrahman and Levy, Omer and Stoyanov, Veselin and Zettlemoyer, Luke},
  booktitle={Proceedings of the 58th annual meeting of the association for computational linguistics},
  pages={7871--7880},
  year={2020}
}

@misc{jetson-nano,
   title = {Jetson-Nano},
   howpublished = {\url{https://developer.nvidia.com/embedded/jetson-nano-developer-kit}},
   year={2019}
}

@article{raffel2020exploring,
  title={Exploring the limits of transfer learning with a unified text-to-text transformer},
  author={Raffel, Colin and Shazeer, Noam and Roberts, Adam and Lee, Katherine and Narang, Sharan and Matena, Michael and Zhou, Yanqi and Li, Wei and Liu, Peter J},
  journal={JMLR},
  volume={21},
  number={140},
  year={2020}
}

@misc{jetson-TX2,
   title = {Jetson-TX2},
   howpublished = {\url{https://developer.nvidia.com/embedded/jetson-tx2}},
   year={2017}
}

@misc{tflite,
  author = {},
   title = {On-device training with tensorflow lite},
   year={2021},
   howpublished = {\url{https://www.tensorflow.org/lite/ examples/on_device_training/overview}}
}

@article{jiang2020mnn,
  title={Mnn: A universal and efficient inference engine},
  author={Jiang, Xiaotang and Wang, Huan and Chen, Yiliu and Wu, Ziqi and Wang, Lichuan and Zou, Bin and Yang, Yafeng and Cui, Zongyang and Cai, Yu and Yu, Tianhang and others},
  journal={Proceedings of MLSys},
  volume={2},
  pages={1--13},
  year={2020}
}

@misc{pytorch,
   title = {PyTorch},
   year={2019},
   howpublished = {\url{https://github.com/pytorch/pytorch}}
}

@inproceedings{ye2022eco,
  title={Eco-FL: Adaptive federated learning with efficient edge collaborative pipeline training},
  author={Ye, Shengyuan and Zeng, Liekang and Wu, Qiong and Luo, Ke and Fang, Qingze and Chen, Xu},
  booktitle={Proceedings of the 51st International Conference on Parallel Processing},
  pages={1--11},
  year={2022}
}

@inproceedings{hao2021eddl,
  title={Eddl: A distributed deep learning system for resource-limited edge computing environment},
  author={Hao, Pengzhan and Zhang, Yifan},
  booktitle={2021 IEEE/ACM Symposium on Edge Computing (SEC)},
  pages={1--13},
  year={2021},
  organization={IEEE}
}

@inproceedings{cai2023efficient,
  title={Efficient federated learning for modern nlp},
  author={Cai, Dongqi and Wu, Yaozong and Wang, Shangguang and Lin, Felix Xiaozhu and Xu, Mengwei},
  booktitle={MobiCom},
  pages={1--16},
  year={2023}
}

@inproceedings{xu2024fwdllm,
  title={$\{$FwdLLM$\}$: Efficient federated finetuning of large language models with perturbed inferences},
  author={Xu, Mengwei and Cai, Dongqi and Wu, Yaozong and Li, Xiang and Wang, Shangguang},
  booktitle={2024 USENIX Annual Technical Conference (USENIX ATC 24)},
  pages={579--596},
  year={2024}
}

@inproceedings{xu2022mandheling,
  title={Mandheling: Mixed-precision on-device dnn training with dsp offloading},
  author={Xu, Daliang and Xu, Mengwei and Wang, Qipeng and Wang, Shangguang and Ma, Yun and Huang, Kang and Huang, Gang and Jin, Xin and Liu, Xuanzhe},
  booktitle={MobiCom},
  pages={214--227},
  year={2022}
}

@inproceedings{gim2022memory,
  title={Memory-efficient dnn training on mobile devices},
  author={Gim, In and Ko, JeongGil},
  booktitle={MobiSys},
  pages={464--476},
  year={2022}
}

@inproceedings{wang2022melon,
  title={Melon: Breaking the memory wall for resource-efficient on-device machine learning},
  author={Wang, Qipeng and Xu, Mengwei and Jin, Chao and Dong, Xinran and Yuan, Jinliang and Jin, Xin and Huang, Gang and Liu, Yunxin and Liu, Xuanzhe},
  booktitle={MobiSys},
  pages={450--463},
  year={2022}
}

@article{lin2022device,
  title={On-device training under 256kb memory},
  author={Lin, Ji and Zhu, Ligeng and Chen, Wei-Ming and Wang, Wei-Chen and Gan, Chuang and Han, Song},
  journal={NeurIPS},
  volume={35},
  year={2022}
}

@article{lester2021power,
  title={The power of scale for parameter-efficient prompt tuning},
  author={Lester, Brian and Al-Rfou, Rami and Constant, Noah},
  journal={arXiv preprint arXiv:2104.08691},
  year={2021}
}

@article{miao2024flexllm,
  title={FlexLLM: A System for Co-Serving Large Language Model Inference and Parameter-Efficient Finetuning},
  author={Miao, Xupeng and Oliaro, Gabriele and Cheng, Xinhao and Wu, Mengdi and Unger, Colin and Jia, Zhihao},
  journal={arXiv:2402.18789},
  year={2024}
}

@inproceedings{houlsby2019parameter,
  title={Parameter-efficient transfer learning for NLP},
  author={Houlsby, Neil and Giurgiu, Andrei and Jastrzebski, Stanislaw and Morrone, Bruna and De Laroussilhe, Quentin and Gesmundo, Andrea and Attariyan, Mona and Gelly, Sylvain},
  booktitle={ICML},
  pages={2790--2799},
  year={2019},
  organization={PMLR}
}

@inproceedings{hu2021lora,
  title={LoRA: Low-Rank Adaptation of Large Language Models},
  author={Hu, Edward J and Wallis, Phillip and Allen-Zhu, Zeyuan and Li, Yuanzhi and Wang, Shean and Wang, Lu and Chen, Weizhu and others},
  booktitle={International Conference on Learning Representations}
}

@article{liu2024tuning,
  title={Y-tuning: An efficient tuning paradigm for large-scale pre-trained models via label representation learning},
  author={Liu, Yitao and An, Chenxin and Qiu, Xipeng},
  journal={Frontiers of Computer Science},
  volume={18},
  number={4},
  pages={184320},
  year={2024},
  publisher={Springer}
}

@article{sung2022lst,
  title={Lst: Ladder side-tuning for parameter and memory efficient transfer learning},
  author={Sung, Yi-Lin and Cho, Jaemin and Bansal, Mohit},
  journal={NeurIPS},
  volume={35},
  year={2022}
}

@inproceedings{yin2023parameter,
  title={Parameter-efficient is not sufficient: Exploring parameter, memory, and time efficient adapter tuning for dense predictions},
  author={Yin, Dongshuo and Han, Xueting and Li, Bin and Feng, Hao and Bai, Jing},
  booktitle={Proceedings of the 32nd ACM International Conference on Multimedia},
  pages={1398--1406},
  year={2024}
}

@inproceedings{patil2022poet,
  title={POET: Training neural networks on tiny devices with integrated rematerialization and paging},
  author={Patil, Shishir G and Jain, Paras and Dutta, Prabal and Stoica, Ion and Gonzalez, Joseph},
  booktitle={International Conference on Machine Learning},
  pages={17573--17583},
  year={2022},
  organization={PMLR}
}

@inproceedings{ouyang2024pluto,
  title={Pluto and Charon: A Time and Memory Efficient Collaborative Edge AI Framework for Personal LLMs Fine-tuning},
  author={Ouyang, Bei and Ye, Shengyuan and Zeng, Liekang and Qian, Tianyi and Li, Jingyi and Chen, Xu},
  booktitle={Proceedings of the 53rd International Conference on Parallel Processing},
  pages={762--771},
  year={2024}
}

@inproceedings{ye2024asteroid,
  title={Asteroid: Resource-Efficient Hybrid Pipeline Parallelism for Collaborative DNN Training on Heterogeneous Edge Devices},
  author={Ye, Shengyuan and Zeng, Liekang and Chu, Xiaowen and Xing, Guoliang and Chen, Xu},
  booktitle={Proceedings of the 30th Annual International Conference on Mobile Computing and Networking},
  pages={312--326},
  year={2024}
}

% \bibliographystyle{IEEEtran}
% \normalem
% \bibliography{reference}

\vspace{-30pt}
\begin{IEEEbiography}
[{\includegraphics[width=1in,height=1.25in,clip,keepaspectratio]{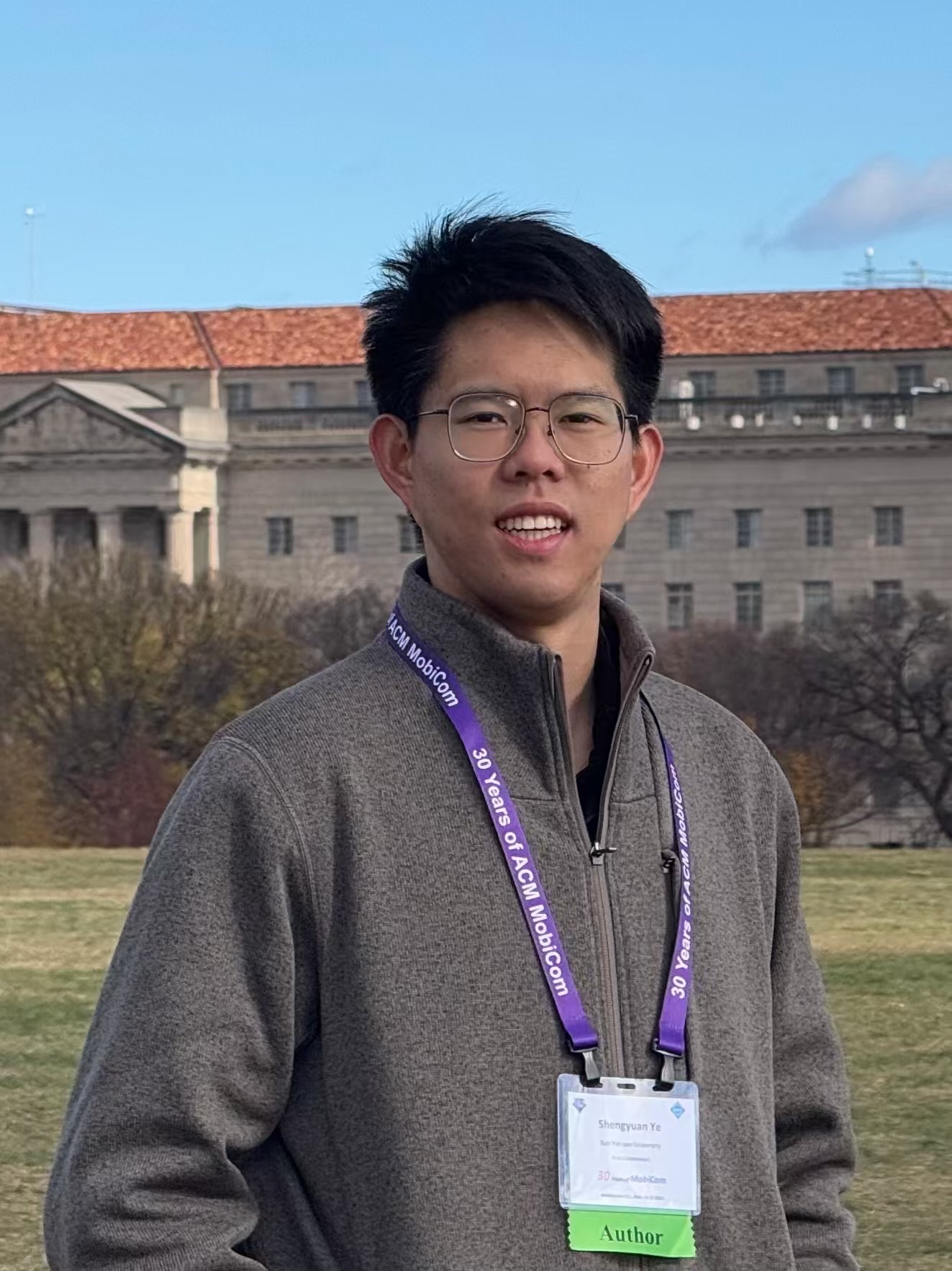}}]{Shengyuan Ye}
received the B.E. degrees from the School of Computer Science and Engineering, Sun Yat-sen University (SYSU), Guangzhou, China, in 2021, where he is currently working toward the Ph.D. degree. His current research interests include mobile edge computing, resource-efficient mobile AI systems, distributed machine learning systems.
\end{IEEEbiography}
\vspace{-30pt}
\begin{IEEEbiography}
[{\includegraphics[width=1in,height=1.25in,clip,keepaspectratio]{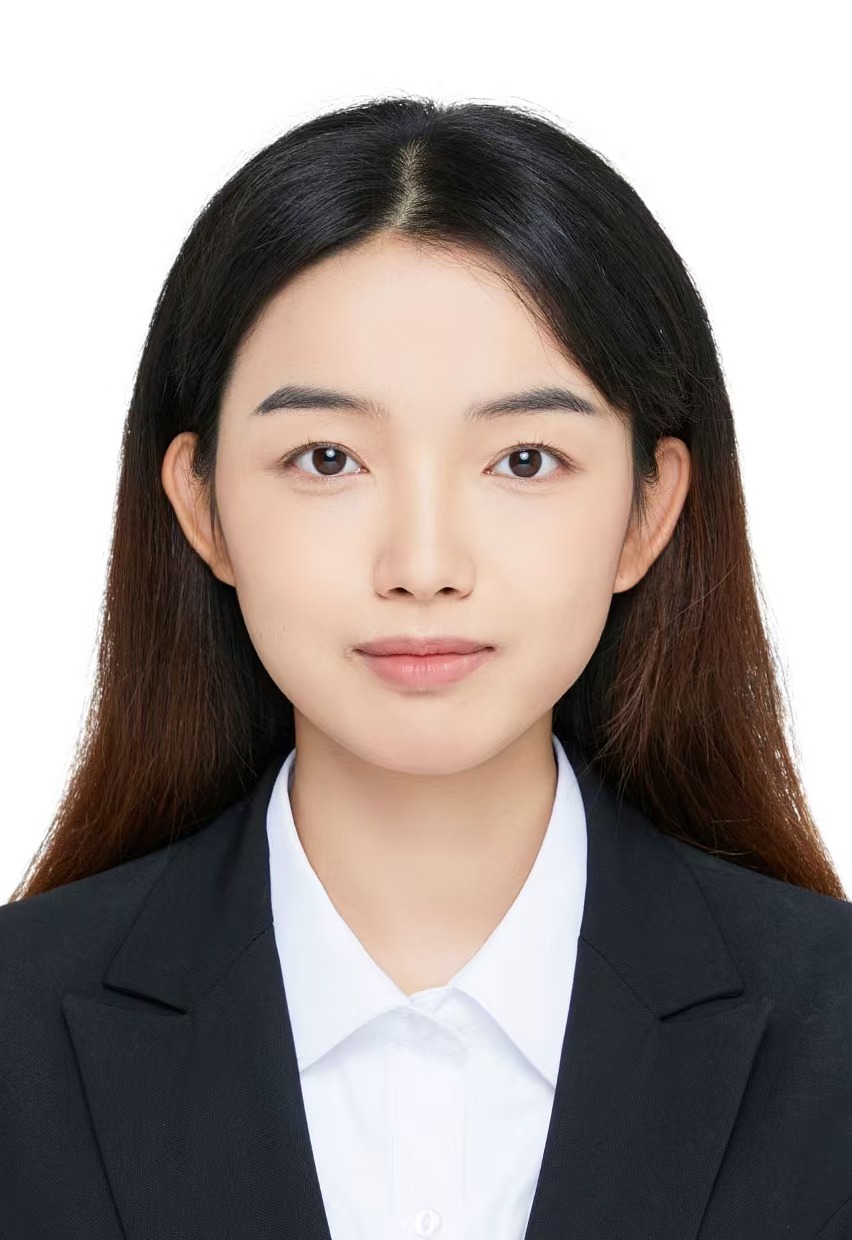}}]{Bei Ouyang}
received the B.S. degree in computer science from the School of Computer Science and Engineering, Sun Yat-sen University (SYSU), Guangzhou, China, in 2023. She is currently pursuing a master's degree at the School of Computer Science and Engineering, Sun Yat-sen University. Her research interests include mobile edge computing and distributed computing.
\end{IEEEbiography}
\vspace{-20pt}
\begin{IEEEbiography}
[{\includegraphics[width=1in,height=1.25in,clip,keepaspectratio]{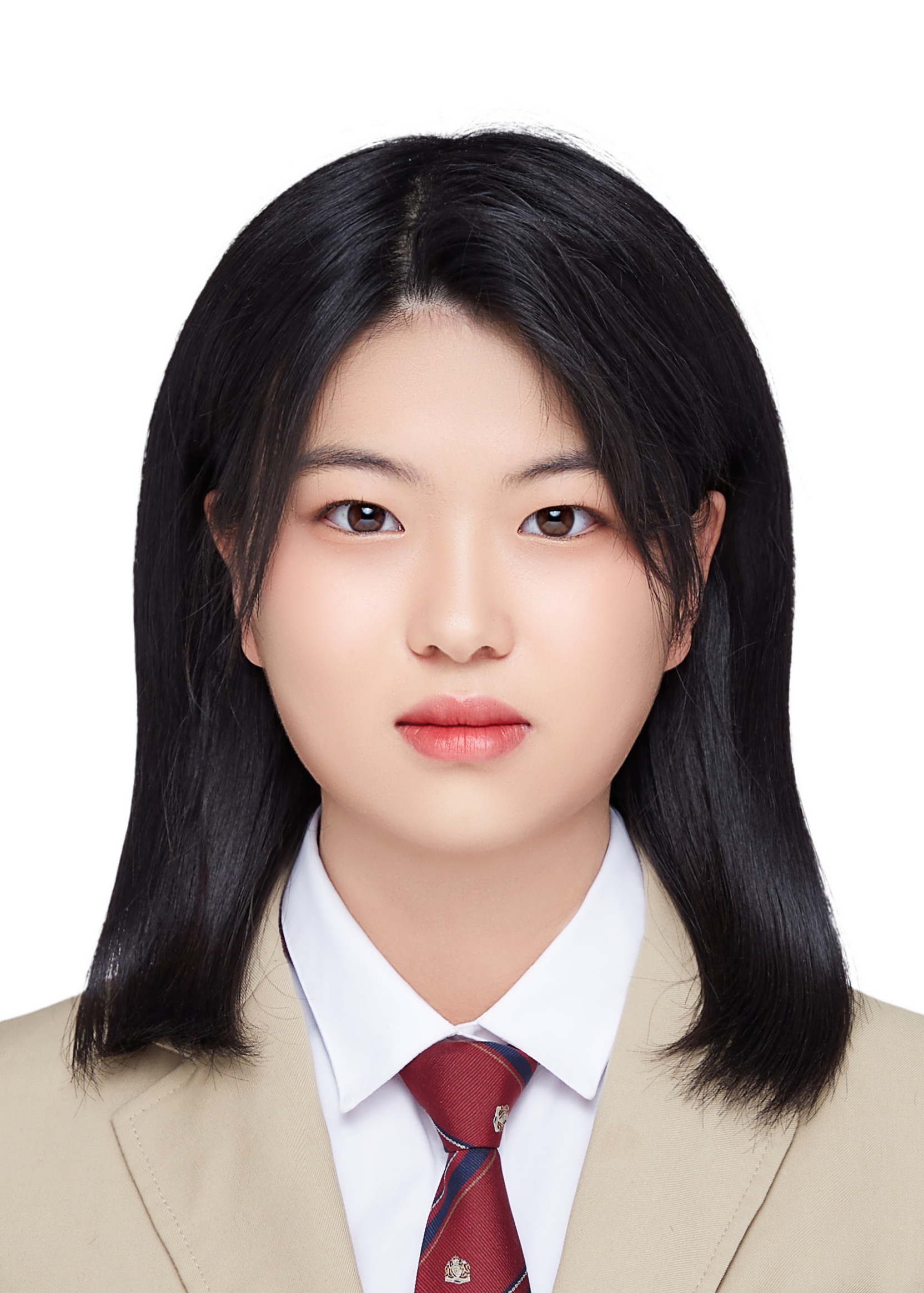}}]{Tianyi Qian}
received the B.S. degree in Computer Science and Technology from the Honors College of Engineering Interdisciplinary Excellence Program, Northwestern Polytechnical University (NPU), Xi'an, China in 2023. She is currently pursuing the master's degree with the School of Computer Science, Sun Yat-sen University (SYSU), Guangzhou, China. Her research interests include mobile edge computing and distributed computing.
\end{IEEEbiography}
\vspace{-20pt}
\begin{IEEEbiography}
[{\includegraphics[width=1in,height=1.25in,clip,keepaspectratio]{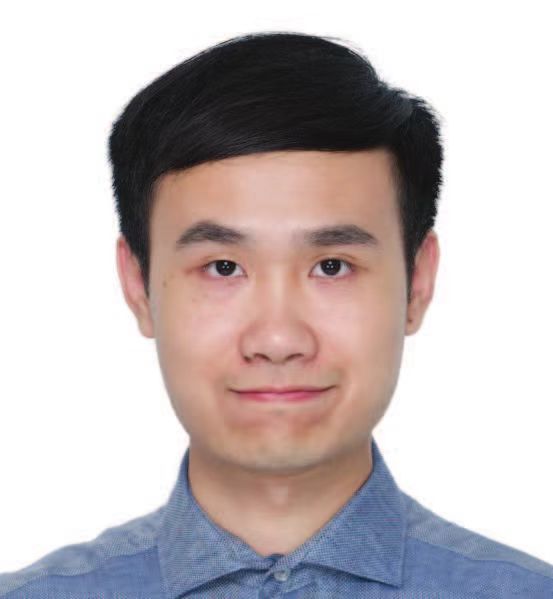}}]{Liekang Zeng}
received the Ph.D. and the B.E. degrees from Sun Yat-sen University, Guangzhou, China.  His current research interests include edge intelligence, mobile computing, and distributed machine learning systems.
\end{IEEEbiography}
\vspace{-20pt}
\begin{IEEEbiography}
[{\includegraphics[width=1in,height=1.25in,clip,keepaspectratio]{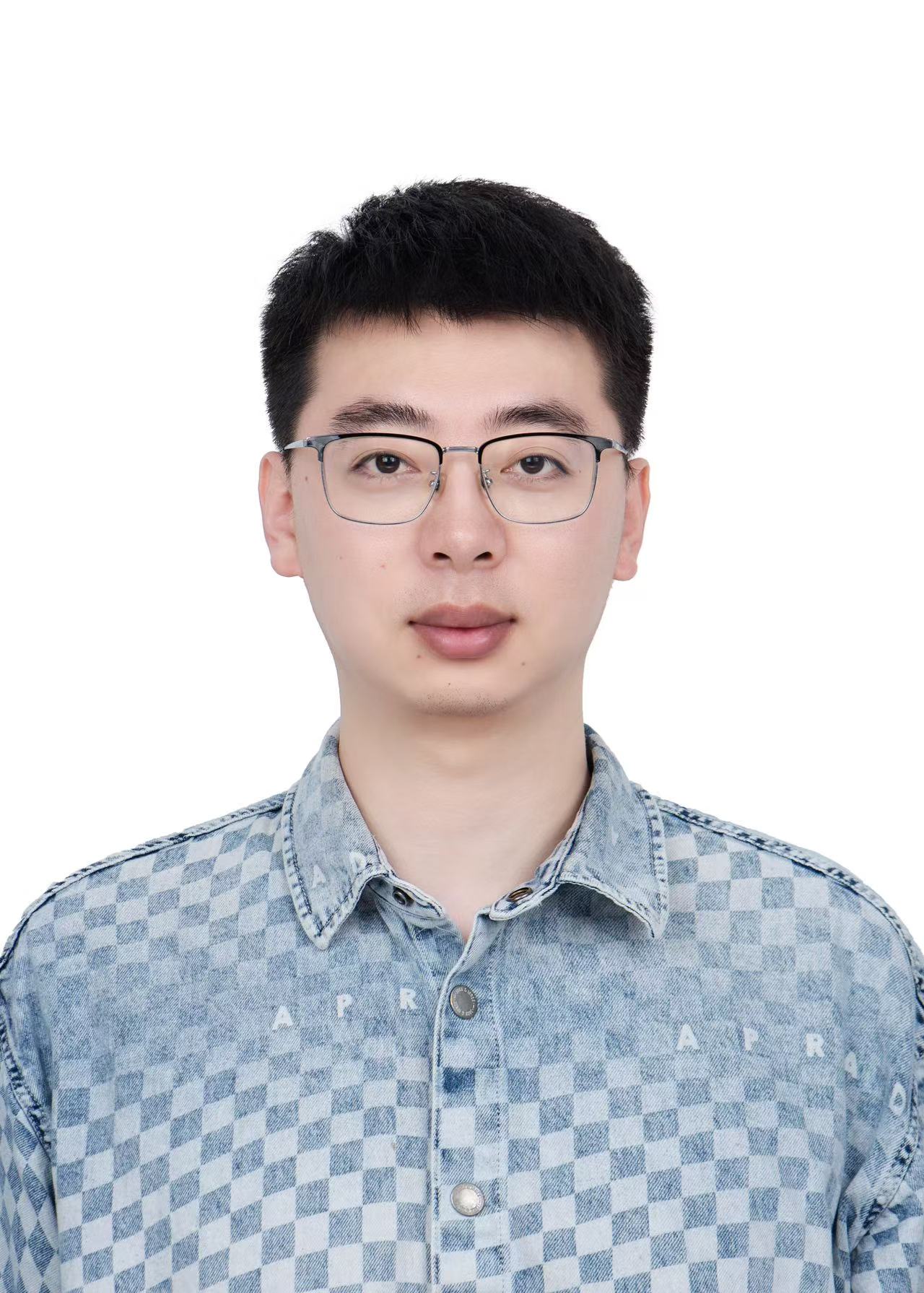}}]{Jingyi Li}
received the B.E. and B.B.A. degrees from South China University of Technology in 2020. He is currently pursuing Ph.D. degree at School of Computer Science and Engineering, Sun Yat-sen University. His research interests include edge intelligence and privacy preservation.
\end{IEEEbiography}
\vspace{-20pt}
\begin{IEEEbiography}
[{\includegraphics[width=1in,height=1.25in,clip,keepaspectratio]{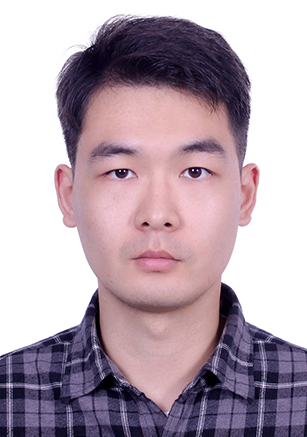}}]{Jiangsu Du}
received the BSc degree from the Wuhan University, in 2016, the MSc degree from the Edinburgh Parallel Computing Center, University of Edinburgh, in 2017, the PhD degree from the School of Computer Science and Engineering, Sun Yat-sen University, in 2022. 
He is now the postdoctoral researcher in the School of Computer Science and Engineering, Sun Yat-sen University. His research interests focus on high performance computing, parallel and distributed artificial intelligent system.
\end{IEEEbiography}
\vspace{-20pt}
\begin{IEEEbiography}
[{\includegraphics[width=1in,height=1.25in,clip,keepaspectratio]{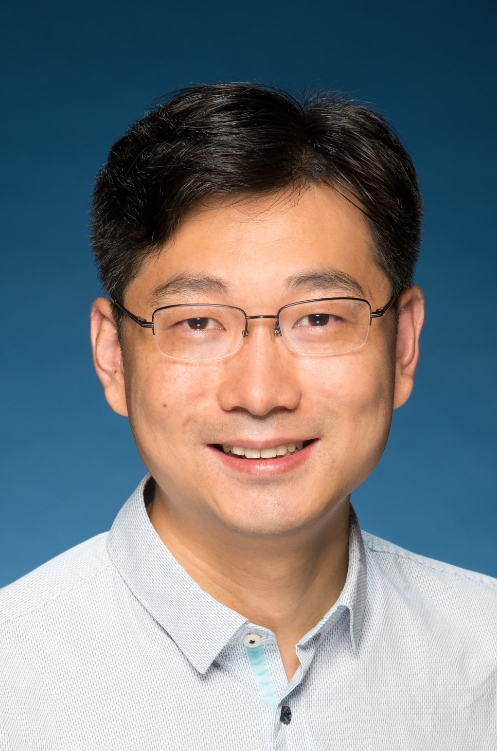}}]{Xiaowen Chu} (Fellow, IEEE) received the B.Eng. degree in computer science from Tsinghua University, Beijing, China, in 1999, and the Ph.D. degree in computer science from The Hong Kong University of Science and Technology, Hong Kong, in 2003. He is currently a Professor with the Data Science and Analytics Thrust, The Hong Kong University of Science and Technology (Guangzhou), Guangzhou, China. His current research interests include Graphics Processing Unit (GPU) computing, distributed machine learning, and wireless networks.
\end{IEEEbiography}
\vspace{-20pt}
\begin{IEEEbiography}
[{\includegraphics[width=1in,height=1.25in,clip,keepaspectratio]{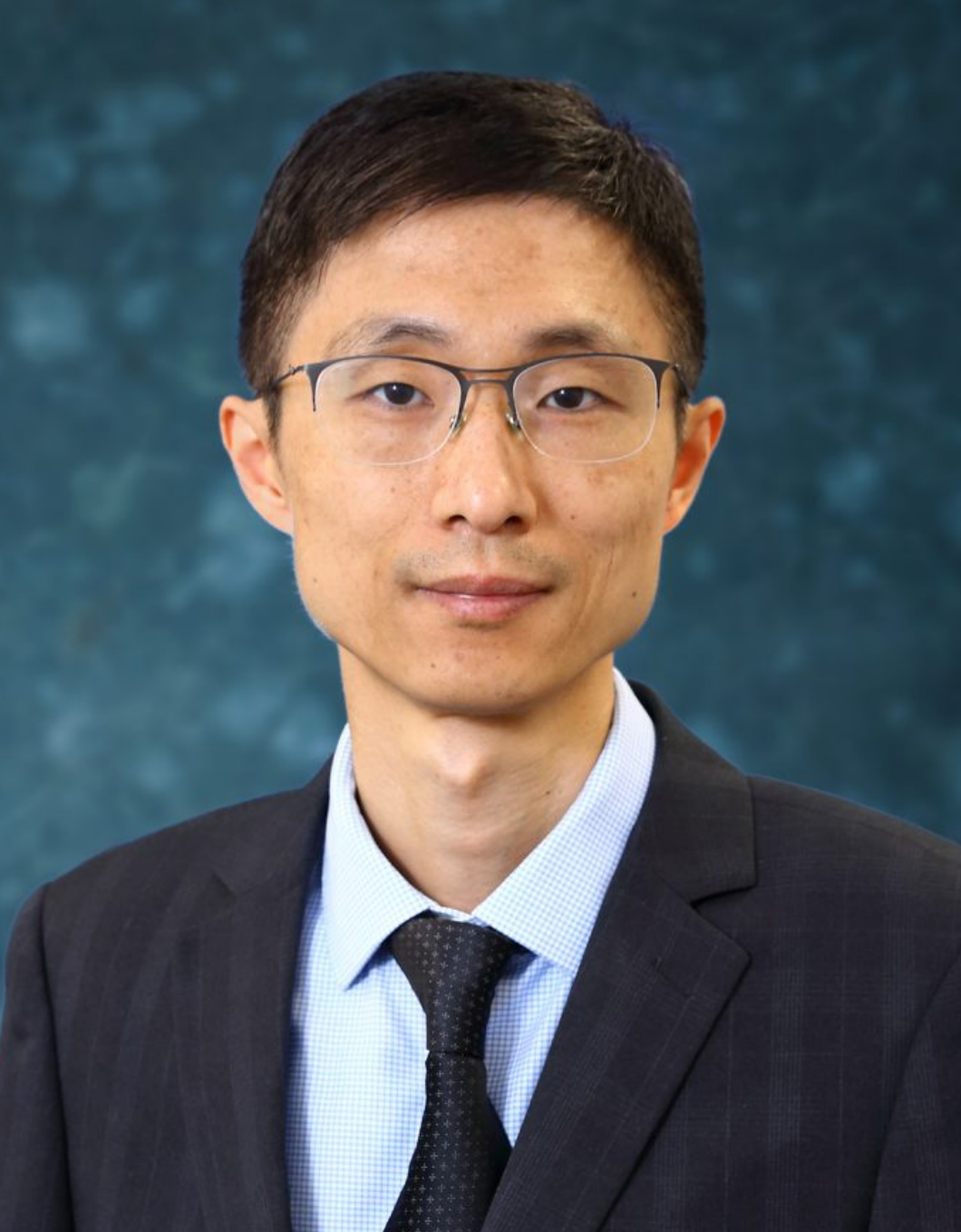}}]{Guoliang Xing} (Fellow, IEEE) received the Ph.D. degree from Washington University, St. Louis, MO, USA, in 2006. He is currently a Professor of information engineering with The Chinese University of Hong Kong, China. Previously, he was a Faculty Member with Michigan State University, East Lansing, MI, USA. His research interests include Internet of Things, smart health, cyber-physical systems, security, and wireless networking. He was the recipient of the three Best Paper Awards and six Best Paper Nominations at leading international conferences. He has led several multi-disciplinary research projects on smart health systems for activity recognition, family wellness, and sleep monitoring. Several mobile health technologies developed in his lab won Best App Awards at the MobiCom conference and were successfully transferred to the industry.
\end{IEEEbiography}
\vspace{-20pt}
\begin{IEEEbiography}
[{\includegraphics[width=1in,height=1.25in,clip,keepaspectratio]{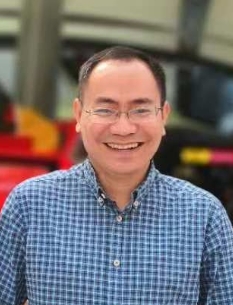}}]{Xu Chen} received the PhD degree in information engineering from the Chinese University of Hong Kong, in 2012. He is a full professor with Sun Yat-sen University, Guangzhou, China, and the vice director of National and Local Joint Engineering Laboratory. His research interests include edge computing, edge intelligence, edge robotics, AI for networking, mobile social networks game theory, deep learning, dynamic optimization, and resource allocation.
\end{IEEEbiography}

\end{document}